\documentclass[reprint,preprintnumbers,nofootinbib,amsmath,amssymb,aps]{revtex4-2}

\usepackage{graphicx}
\usepackage{dcolumn}
\usepackage{bm}

\usepackage{color}
\usepackage[dvipsnames, table]{xcolor}
\definecolor{blue}{rgb}{0,0,0.5}
\usepackage[colorlinks,linkcolor=blue,urlcolor=blue,citecolor=blue]{hyperref}
\usepackage{graphicx}
\usepackage{slashed}
\usepackage{enumitem}
\usepackage{bm}
\usepackage{physics}
\usepackage{xspace}
\usepackage[T1]{fontenc}
\usepackage{lmodern}
   
\usepackage{mathrsfs}

\newcommand{\be}{\begin{equation}}
\newcommand{\ee}{\end{equation}}
\newcommand{\bea}{\begin{eqnarray}}
\newcommand{\eea}{\end{eqnarray}}

\newcommand{\mc}{\mathcal}

\newcommand{\nn}{\nonumber}
\newcommand{\noi}{\noindent}

\newcommand{\miss}{{\rm miss}}
\newcommand{\Emiss}{E_{\miss}}
\newcommand{\Mrecoil}{M_{\rm recoil}}

\def\T{{\rm T}}

\def\mt{M_{\rm T}}
\def\MT2{M_{\rm T2}}
\def\M2{M_{\rm 2}}
\def\k#1{{\boldsymbol{k}}_{#1}}

\def\p#1{{\boldsymbol{p}}_{#1}}

\def\maos{{\rm maos}}

\def\max{{\rm max}}
\def\min{{\rm min}}

\def\RpT{R_{p_\T}}
\def\RkT{R_{k_\T}}

\def\xip{\xi_{p}}
\def\xik{\xi_{k}}
\newcommand{\tauRF}{\tau\mbox{-}{\rm RF}}
\newcommand{\ma}{m_\phi}
\newcommand{\MeV}{{\rm MeV}}
\newcommand{\GeV}{{\rm GeV}}
\newcommand{\oxt}{1\!\times\!3}
\newcommand{\oxo}{1\!\times\!1}

\DeclareOldFontCommand{\rm}{\normalfont\rmfamily}{\mathrm}
\DeclareOldFontCommand{\sf}{\normalfont\sffamily}{\mathsf}
\DeclareOldFontCommand{\tt}{\normalfont\ttfamily}{\mathtt}
\DeclareOldFontCommand{\bf}{\normalfont\bfseries}{\mathbf}
\DeclareOldFontCommand{\it}{\normalfont\itshape}{\mathit}
\DeclareOldFontCommand{\sl}{\normalfont\slshape}{\@nomath\sl}
\DeclareOldFontCommand{\sc}{\normalfont\scshape}{\@nomath\sc}

\begin{document}

\preprint{CERN-TH-2021-101, LAPTH-023/21, CTPU-PTC-21-28}

\title{\boldmath $\tau \to \ell +$ invisible\\through invisible-savvy collider variables}

\author{Diego Guadagnoli}
\email{diego.guadagnoli@lapth.cnrs.fr}
\affiliation{%
Theoretical Physics Department, CERN, 1211 Geneva 23, Switzerland\\
LAPTh, CNRS et Universit\'{e} Savoie Mont-Blanc, F-74941 Annecy, France
}%

\author{Chan Beom Park}
\email{cbpark@ibs.re.kr}
\affiliation{%
Center for Theoretical Physics of the Universe, Institute for Basic Science (IBS), Daejeon, 34126, Korea
}%

\author{Francesco Tenchini}
\email{francesco.tenchini@desy.de}
\affiliation{%
Dipartimento di Fisica, Universit\`a di Pisa and INFN Sezione di Pisa, I-56127 Pisa, Italy\\
Deutsches Elektronen-Synchrotron (DESY), Hamburg, D-22607, Germany
}%

\begin{abstract}
\noi New particles $\phi$ in the MeV-GeV range produced at colliders and escaping detection can be searched for at operating $b-$ and $\tau-$factories such as Belle II. A typical search topology involves pair-produced $\tau$s (or mesons), one of which decaying to visibles plus the $\phi$, and the other providing a tag.
One crucial impediment of these searches is the limited ability to reconstruct the parents' separate boosts. This is the case in the `typical' topology where both decay branches include escaping particles.
We observe that such topology lends itself to the use of kinematic variables such as $M_2$, designed for pairwise decays to visibles plus escaping particles, and endowed with a built-in (`MAOS') way to efficiently guess the parents' separate boosts. Starting from this observation, we construct several kinematic quantities able to discriminate signal from background, and apply them to a benchmark search, $\tau \to e + \phi$, where $\phi$ can be either an axion-like particle or a hidden photon. Our considered variables can be applied to a wider range of topologies than the current reference technique, based on the event thrust, with which they are nearly uncorrelated.
Application of our strategy leads to an improvement by a factor close to 3 in the branching-ratio upper limit for $\tau \to e \phi$, with respect to the currently expected limit, assuming $m_\phi \lesssim 1$ MeV. For example, we anticipate a sensitivity of $1.7 \times 10^{-5}$ with the data collected before the 2022 shutdown.
\end{abstract}

\maketitle

\noi New light particles are commonplace in Standard-Model (SM) extensions and allow to elegantly solve problems of both conceptual and observational nature, see e.g.~\cite{Beacham:2019nyx}. Depending on the couplings structure, these particles may actually be as heavy as a GeV. Remarkably, new scalars in the MeV-GeV range with larger than weak couplings to SM matter are fully compatible with the body of knowledge we have on stable matter~\cite{Lanfranchi:2020crw}, because most constraints, notably astrophysical data, apply to interactions with 1$^{\rm st}$-generation matter only. Theoretically, there is no compelling reason why these particles' couplings should be universal across generations, or flavour-diagonal~\cite{Georgi:1986df}. Meson or $\tau$ decays at colliders are especially suited to test such couplings, not only by definition, but also because of the large statistics and accuracy now attainable.

A common hypothesis that allows for minimal model dependence is that these light particles, once produced in the decay, escape detection. The strongest limits are obtained in missing-energy searches where the new particle is either produced in the beam interaction with a fixed target, or is the product of collisions whose initial state is very well known -- e.g. $e^+ e^-$ beams. For a recent comprehensive review see Ref.~\cite{Lanfranchi:2020crw}.

A prototype example of search under the above hypothesis is $\tau \to \ell$ ($= e$ or $\mu$) plus an axion-like particle (ALP, denoted by $\phi$) \cite{Calibbi:2020jvd}, performed at Mark~III~\cite{Baltrusaitis:1985fh} and ARGUS~\cite{Albrecht:1995ht}, and on-going at Belle II~\cite{Tenchini:2020njf}.%
\footnote{A recent limit was also placed by \cite{Bryman:2021rtr}.}%
At these facilities, the parent $\tau$'s are pair-produced at well-defined energies and their decay products are collected over a large angular acceptance, which allows for an accurate estimate of the total missing energy of the system. The dominant background to this kind of search is represented by SM processes also containing undetected particles, notably neutrinos. Hence, for the rare signal and the overwhelming background alike, the separate momenta of the pair-produced $\tau$'s are unknown. To pinpoint the signal, the reference strategy has historically been to estimate the signal-$\tau$ momentum using the visible momenta on the tag side. If for the latter one assumes $\tau \to 3\pi \nu$ (note that the $3\pi$ allow for a high-quality vertex), the signal-$\tau$ momentum may then be estimated via the relations $\widehat{\vb*{p}}_{\tau} \approx - \sum_i^{\rm tag} \widehat{\vb*{p}}_{\pi_i}$, $E_{\tau} \approx \sqrt s / 2$~\cite{Albrecht:1995ht}.\footnote{Here and henceforth, a hat denotes a unit vector.}

A generalisation of this technique \cite{Abudinen:2020das}, representing the current state-of-the-art, takes advantage of the `thrust axis' \cite{Brandt:1964sa,Farhi:1977sg} of the event, identified from the maximum of the `thrust scalar' $T \equiv \sum_i \p{i} \cdot \widehat{\vb*{n}} / |\p{i}|$, where $\p{i}$ denotes {\em all} the visible momenta in the decay. This direction can be used for approximating the signal-$\tau$ momentum~\cite{Tenchini:2020njf} as
\begin{equation}
  p_{\tau} = \frac{\sqrt{s}}{2} \big( 1, \, \widehat{\vb*{n}} \, \sqrt{1 - 4 m_\tau^2 / s} \big)~.
\end{equation}
In either case, the spectrum of the signal-side daughter lepton accompanying the $\phi$ particle is then calculated through a boost to the rest frame of the parent $\tau$.

We propose a different approach based on the following observations:
{\em (i)} signal and background decays have a common topology, consisting of visible final states plus either the elusive $\phi$ (signal), or neutrinos (backgrounds). There exists an arsenal of kinematic variables that are designed precisely for such pairwise decay topology, in particular the `stransverse mass' $\MT2$ \cite{Lester:1999tx,Barr:2003rg} and Lorentz-invariant generalisations thereof \cite{Barr:2011xt}, here collectively denoted as $\M2$; 
{\em (ii)} $\M2$ is a minimisation procedure in the three unknowns constituting the invisible 3-momentum on one of the two decay branches.%
\footnote{At lepton colliders, barring initial-state radiation, the invisible 3-momentum on the other branch is known, because the total missing momentum is.}
The minimum, henceforth referred to as {\em $\M2$-Assisted On-Shell}, or `MAOS', invisible momentum \cite{Cho:2008tj,Park:2011uz}, is distributed around the corresponding {\em true} invisible 3-momentum. $\M2$ thereby offers a `built-in' estimator of the invisible 3-momenta, separately for the two branches -- which addresses the underlying challenge of the search; 
{\em (iii)} We expect $\M2$, as well as MAOS momenta, to show negligible correlation with variables built out of visible momenta only. Because of this small correlation all of these variables can be profitably combined. Our results show that the performance of such combination is substantially higher than the case where the different variables discussed are used individually.

We introduce our idea in the context of $\tau \to \ell + \phi$,\footnote{In all formulae it is understood that $\ell = e,\mu$. All numerics will assume $\ell = e$ for consistency with the analysis in \cite{Tenchini:2020njf}.} with $\phi$ an ALP. However, the kinematic methods discussed may be applied to any other beyond-SM scenario with the same topology, e.g. lepton flavour violating couplings mediated by a spin-0 or spin-1 particle. We will make comments notably on the case of a `hidden photon'.

The signal of interest is
\be
\label{eq:S_1x3}
  e^+ e^- \to \tau (\to \ell \phi) ~\tau (\to \rm{3\pi \nu})\,,
\ee
where we have omitted charge specifications on the r.h.s., and we assume that the $\phi$ escapes undetected. The $\tau$ decay to three charged pions is used as a tag. We denote this channel as `$\oxt$'. The dominant, irreducible background is
\be
\label{eq:B_1x3}
  e^+ e^- \to \tau (\to \ell \nu \bar \nu) ~\tau (\to 3\pi \nu)\,,
\ee
while other channels such as $\tau (\to \pi \nu) ~\tau (\to 3\pi \nu)$ are all suppressed by either PID requirements, kinematic selections, event-shape analysis or track vertexing. The tag decay in eq. (\ref{eq:S_1x3}) is chosen for comparison with current state-of-the-art measurements; however, MAOS momenta can also be calculated for 1-prong tags such as $\tau (\to \ell \nu \bar \nu)$, for which the ARGUS/thrust method is not available. We will thus also consider this `$\oxo$' channel, whose irreducible background is $\tau (\to \ell \nu \bar \nu) \tau (\to \ell \nu \bar \nu)$.

Clearly, in terms of the decay topology the signal and backgrounds thus differ only by the number of invisible particles. %
As mentioned, this topology lends itself to the use of $\MT2$ and its generalizations. Such variables have been extensively applied to high-$\vb*{p}_\T$ searches, notably 
of pair-produced supersymmetric particles. Instead, to our knowledge, this approach has
never been considered for decays of pair-produced mesons or leptons, of the kind 
of interest here.

The $\MT2$ variable is the two-decay-chain generalization of the $\mt$ variable \cite{Smith:1983aa,Barger:1983wf}, used since CERN's UA1 experiment to measure the $W$ mass in $W \to \ell \nu$, and defined from the inequality $m_W^2 \ge m_\ell^2 + m_\nu^2 + 2 (E^\ell_{\T} E^\nu_{\T} - \p{\T}^\ell \p{\T}^\nu) ~\equiv~ \mt^2$
implying that the $\mt$ endpoint allows to measure $m_W$. If one has {\em two} parents decaying to visible products with collective momenta $\p{1,2}$ for the two branches, plus invisible final states on either branch, one may generalise the above argument with $\max\{\mt(\mbox{branch}_1), \mt(\mbox{branch}_2)\}$ -- the largest $\mt$ will furnish the best lower bound. While the two invisible momenta, $\k{1,2}$, are not known individually, they fulfil the constraint $\k{1\T} + \k{2\T} = \vb*P_{\T}^{\miss}$, where the r.h.s. denotes the measured transverse momentum imbalance. Hence, the `most conservative max' one can take is the minimum over the configurations fulfilling such constraints, 
i.e.
\begin{align}
\label{eq:MT2def}
  \MT2 =
  & \min_{\vb*{k}_{1\T}, \, \vb*{k}_{2\T}}
    \Big[ \max \Big\{ \mt(\p{1\T}, \, \k{1\T}), \, \mt(\p{2\T}, \,
    \k{2\T})
    \Big\} \Big]  \nonumber\\
  & \text{subject to}\,\, \vb*{k}_{1\T} + \vb*{k}_{2\T} =
    \vb*{P}_{\T}^\text{miss} ,
\end{align}
which {\em defines} $\MT2$ \cite{Lester:1999tx,Barr:2003rg}.

The subscript ``\textit{\T}'' in the above discussion denotes the projection onto 
the plane transverse to the beam direction. Such projection is unnecessary 
at lepton colliders like Belle~II, where the full, transverse as well as longitudinal, momentum imbalance is reconstructible.
We can then employ the fully Lorentz-invariant extension of $\MT2$, known as $\M2$~\cite{Barr:2011xt} (see also \cite{Ross:2007rm,Cho:2014naa}).
As elucidated in Ref. \cite{Cho:2014naa}, several variations of the $\M2$ variable can actually be used for one and the same topology, depending on the kinematic constraints --~i.e. on-shell mass relations~-- that are imposed in the minimisation and those that are not. This feature makes $\M2$ an extremely versatile tool. Since the center-of-mass system (CMS) energy is fixed, we focus on the following definition
\begin{align}
\label{eq:M2_ours}
  \M2 =
  & \min_{\vb*{k}_{1}, \, \vb*{k}_{2}}
    \Big[ \max \Big\{ M(p_{1}, \, k_{1}), \, M(p_{2}, \, k_{2})
    \Big\} \Big]  \nonumber\\
  & \text{subject to}\,\,
    \begin{cases}
      \vb*{k}_{1} + \vb*{k}_{2} =
      \vb*{P}^\text{miss} , \\
      (p_1 + p_2 + k_1 + k_2)^2 = s ,
    \end{cases}
\end{align}%
where $s$ is the squared collision energy, and $p_i (k_i)$ denote here and henceforth the visible- (invisible-)system total momenta on the decay branch $i = 1, 2$. One may further subject eq. (\ref{eq:M2_ours}) to the decaying parents' on-shell mass relations, i.e. $(p_1 + k_1)^2 = (p_2 + k_2)^2 = m_\tau^2$. However, the additional constraints lead to the same minimum as our $M_2$ definition \cite{Konar:2015hea}.%
\footnote{In the definition (\ref{eq:M2_ours}), $\M2$ is similar to $M_{2 {\rm Cons}}$ devised to study
$H \to \tau^+ \tau^-$ at hadron colliders~\cite{Konar:2015hea, Konar:2016wbh}, except that the constraint on the missing longitudinal momentum is not adopted.}
\begin{figure}[t]
  \begin{center}
    \includegraphics[width=0.235\textwidth]{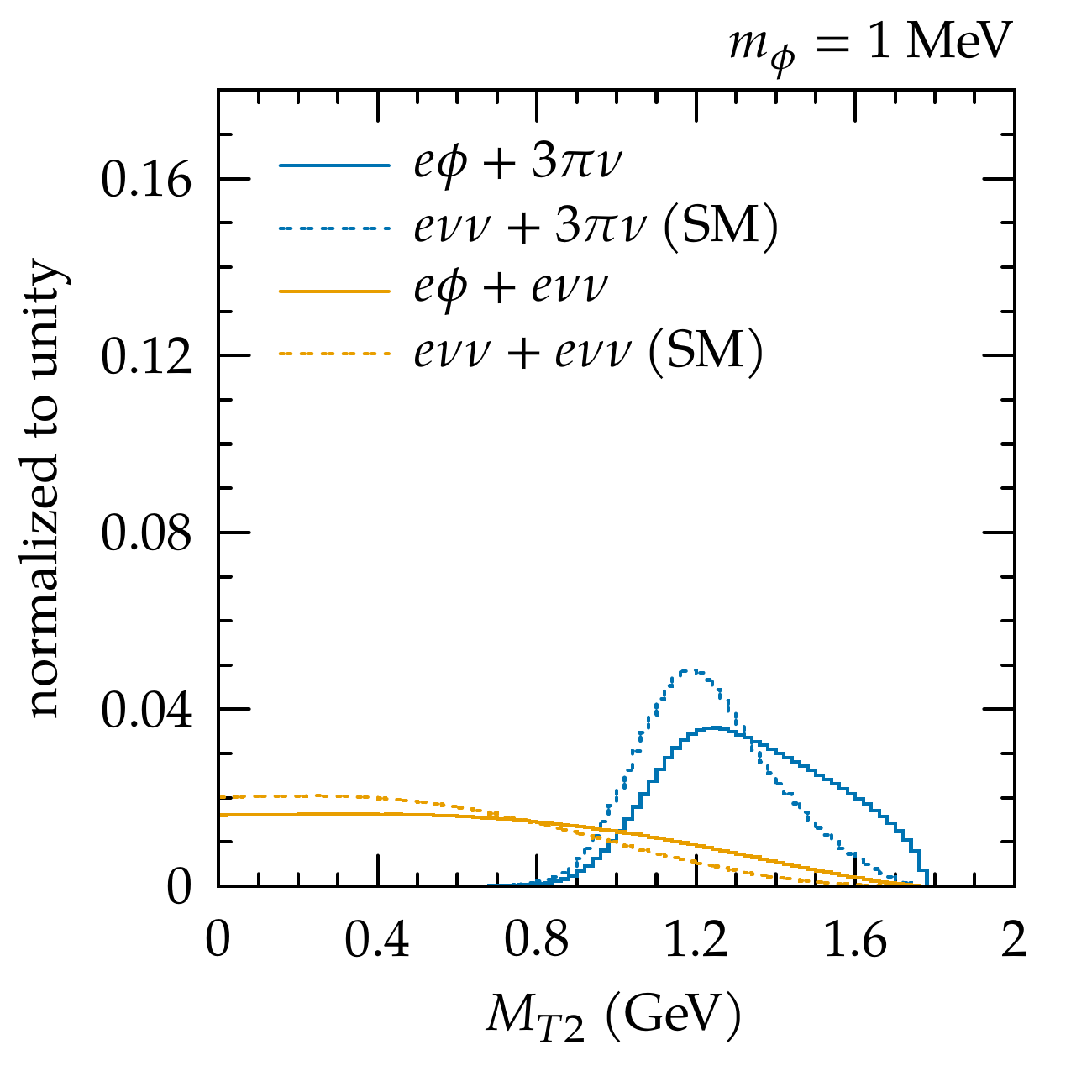}
    \includegraphics[width=0.235\textwidth]{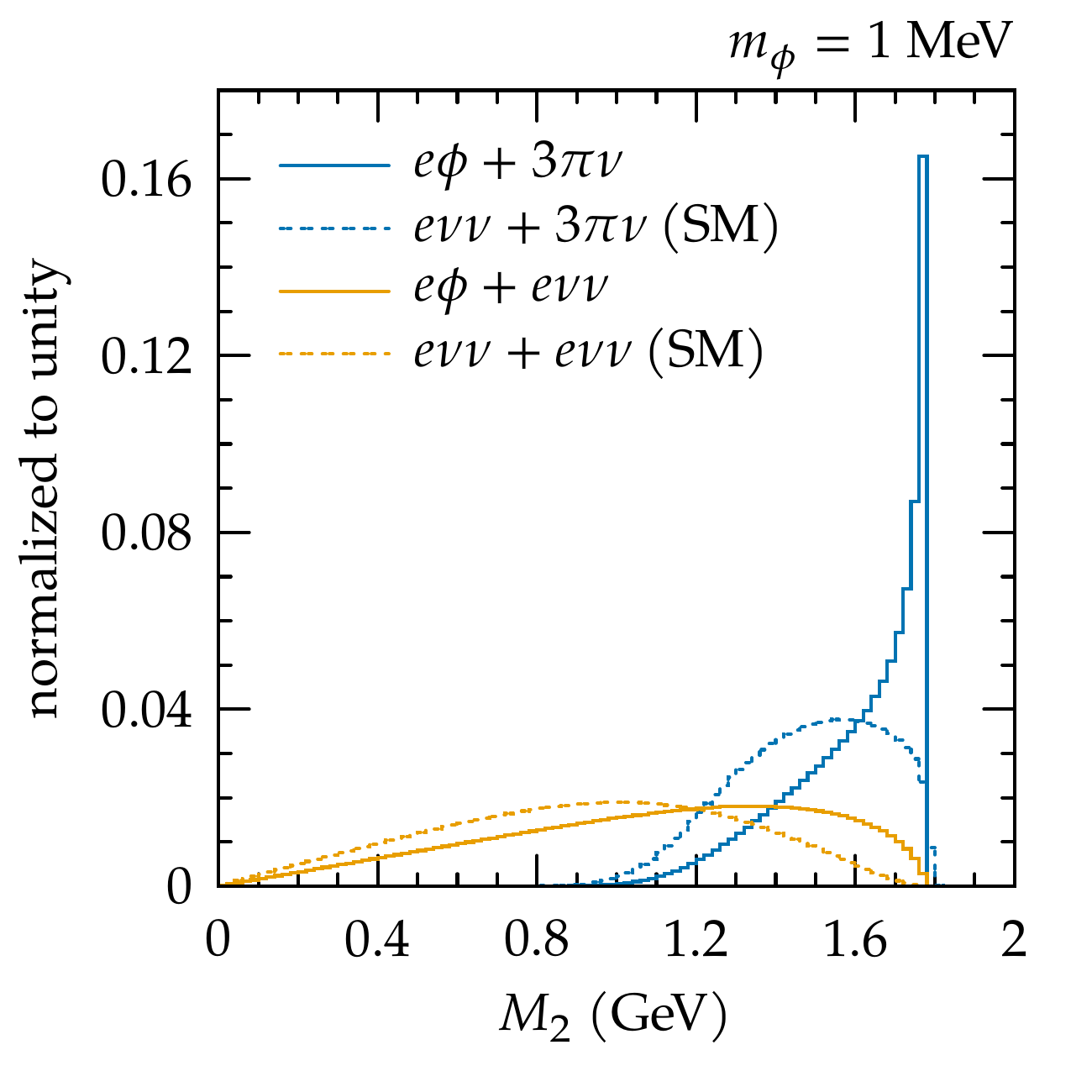}
  \end{center}
  \caption{\label{fig:MT2_vs_M2}
Comparison between $\MT2$ and $\M2$ in the separation of signals and backgrounds (full vs. dashed lines) for the $\oxt$ (blue) and the $\oxo$ (orange) topologies. We show the case $\ma = 1$~MeV.
}
\end{figure}%

\begin{figure}[t]
  \begin{center}
    \includegraphics[width=0.23\textwidth]{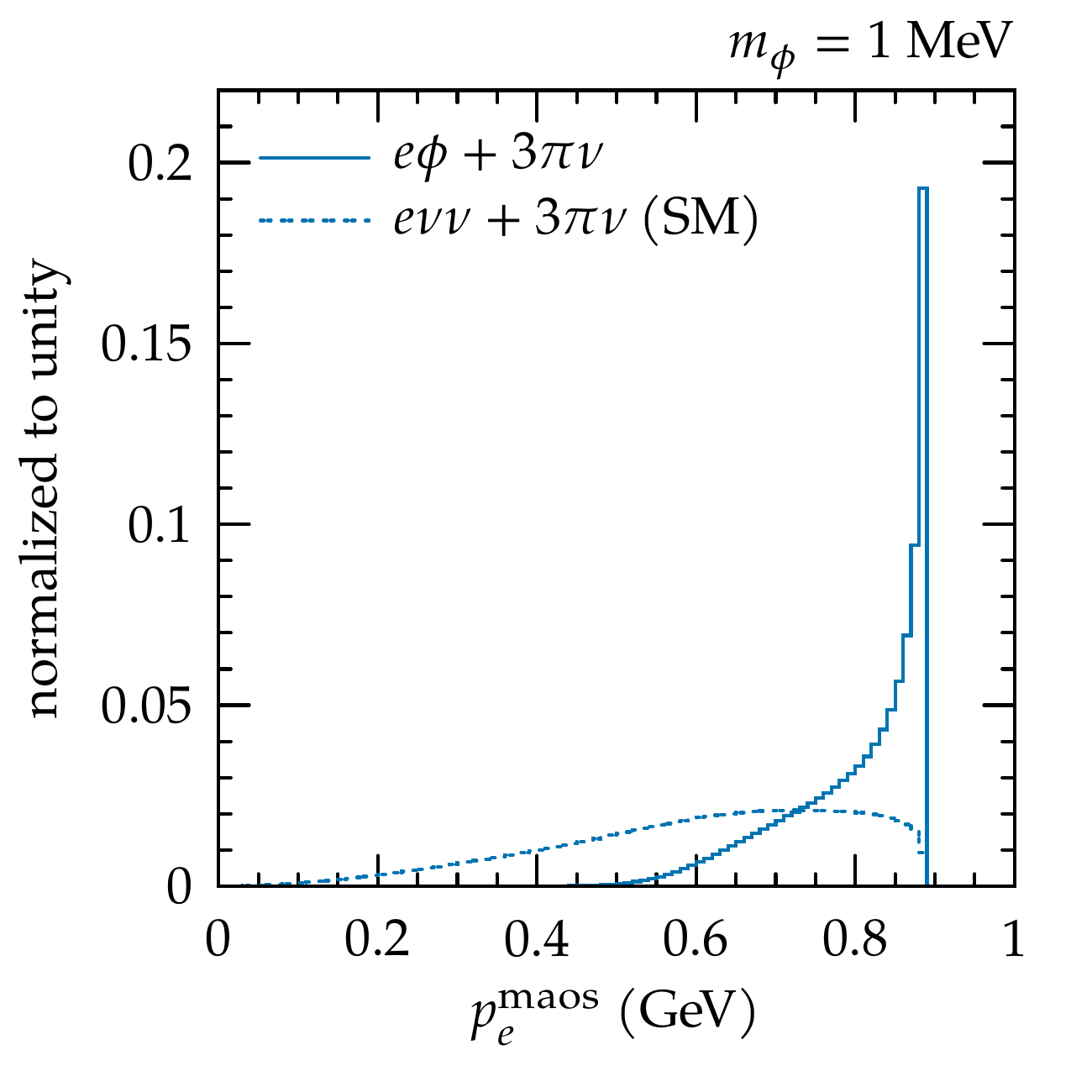}
    \includegraphics[width=0.23\textwidth]{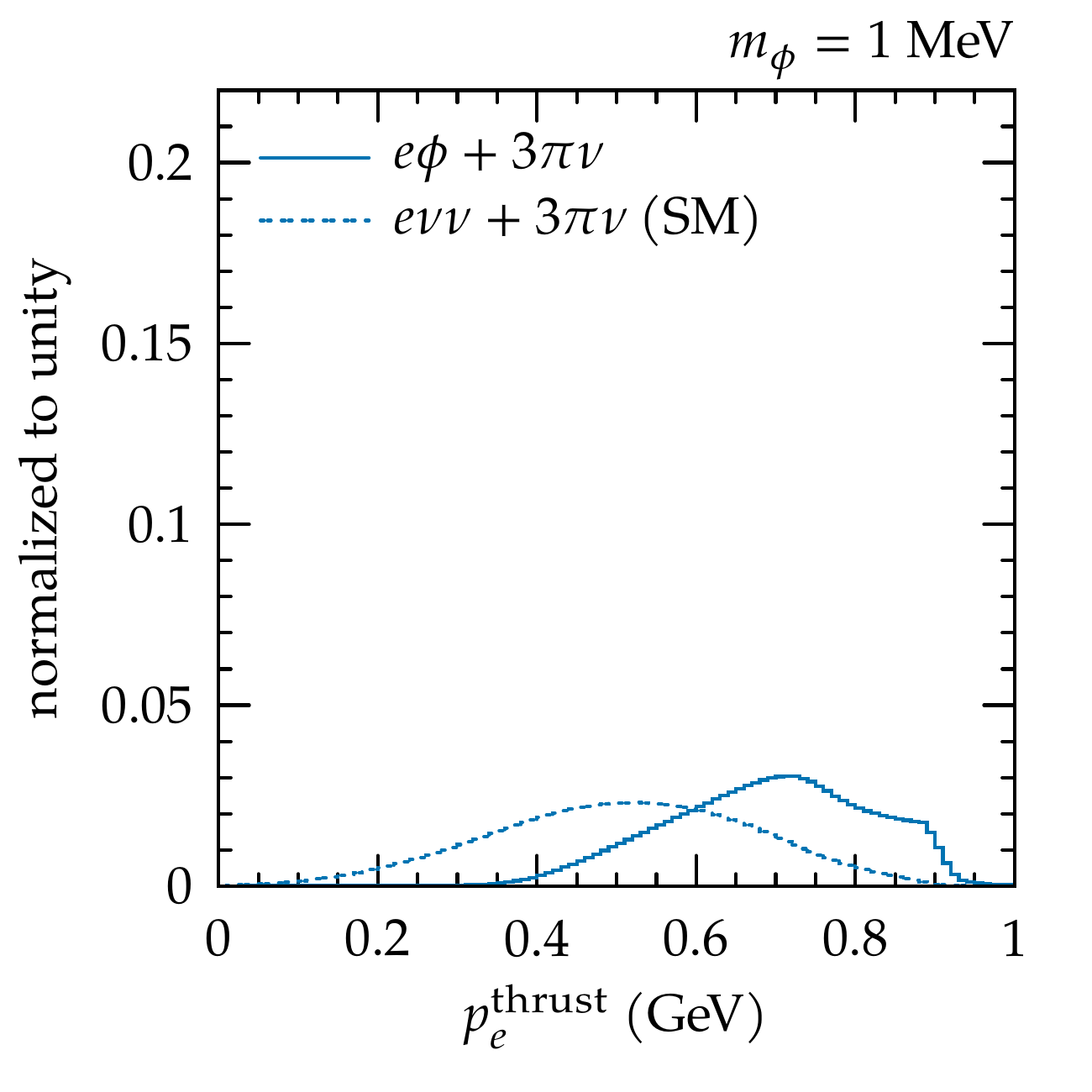}
  \end{center}
  \caption{\label{fig:pe}
Distribution for $p_e^{\maos} \equiv |\vb*{p}_e|^{\maos}_{{\tauRF}}$ and for the corresponding quantity obtained through the thrust, for the $\oxt$ channel and $\ma = 1$ MeV.
}
\end{figure}

\begin{figure}[t]
  \begin{center}
    \includegraphics[width=0.235\textwidth]{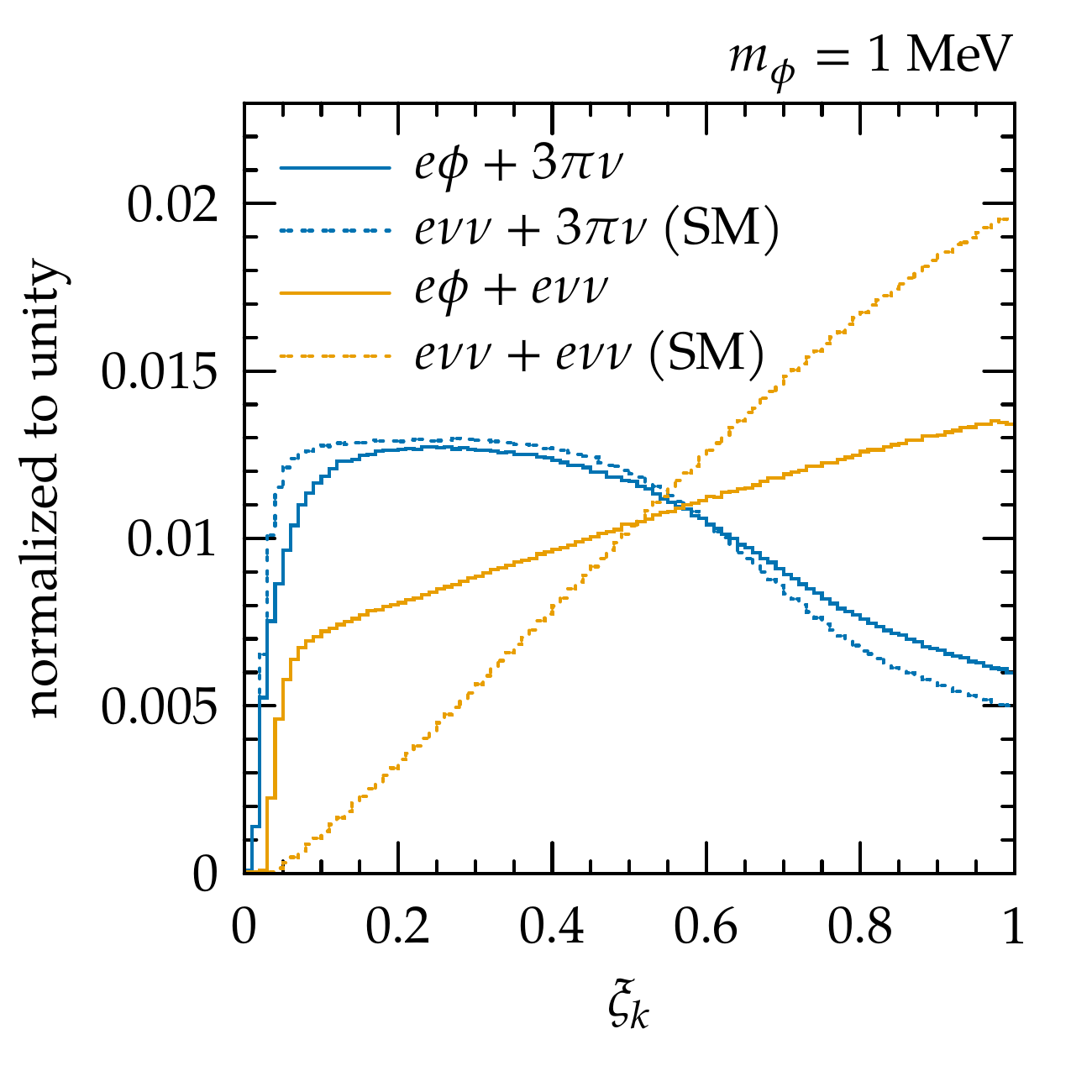}
    \includegraphics[width=0.235\textwidth]{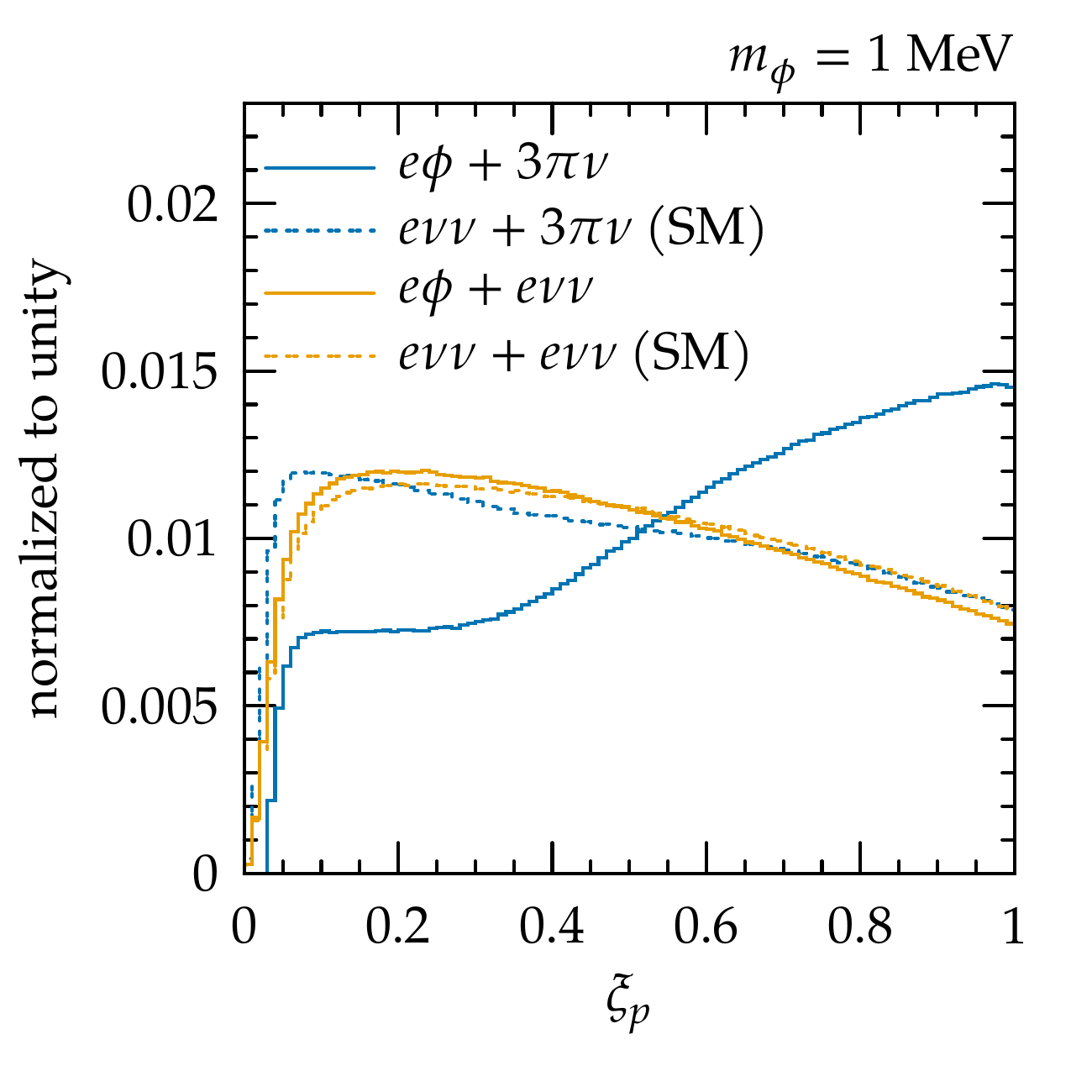}
    \includegraphics[width=0.235\textwidth]{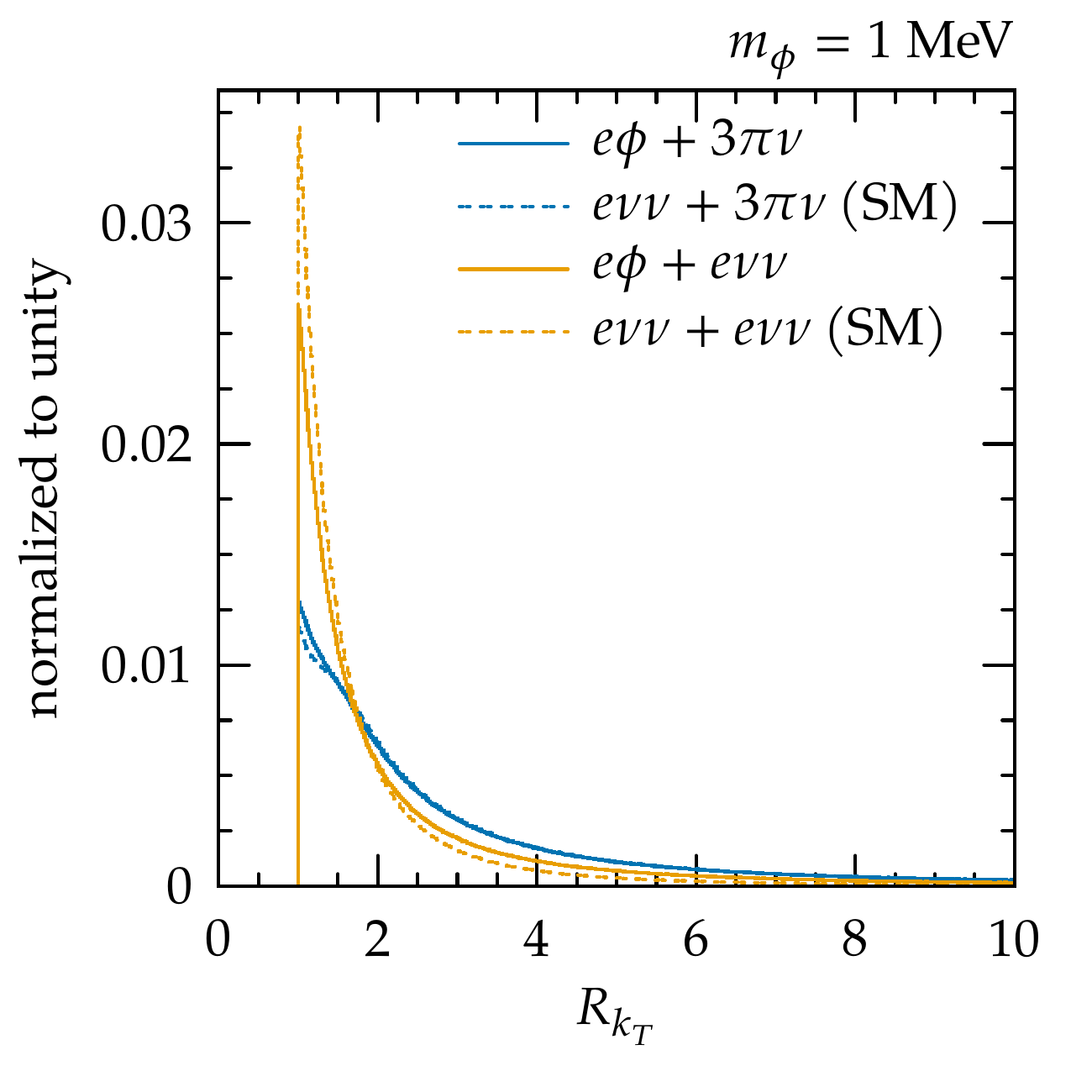}
    \includegraphics[width=0.235\textwidth]{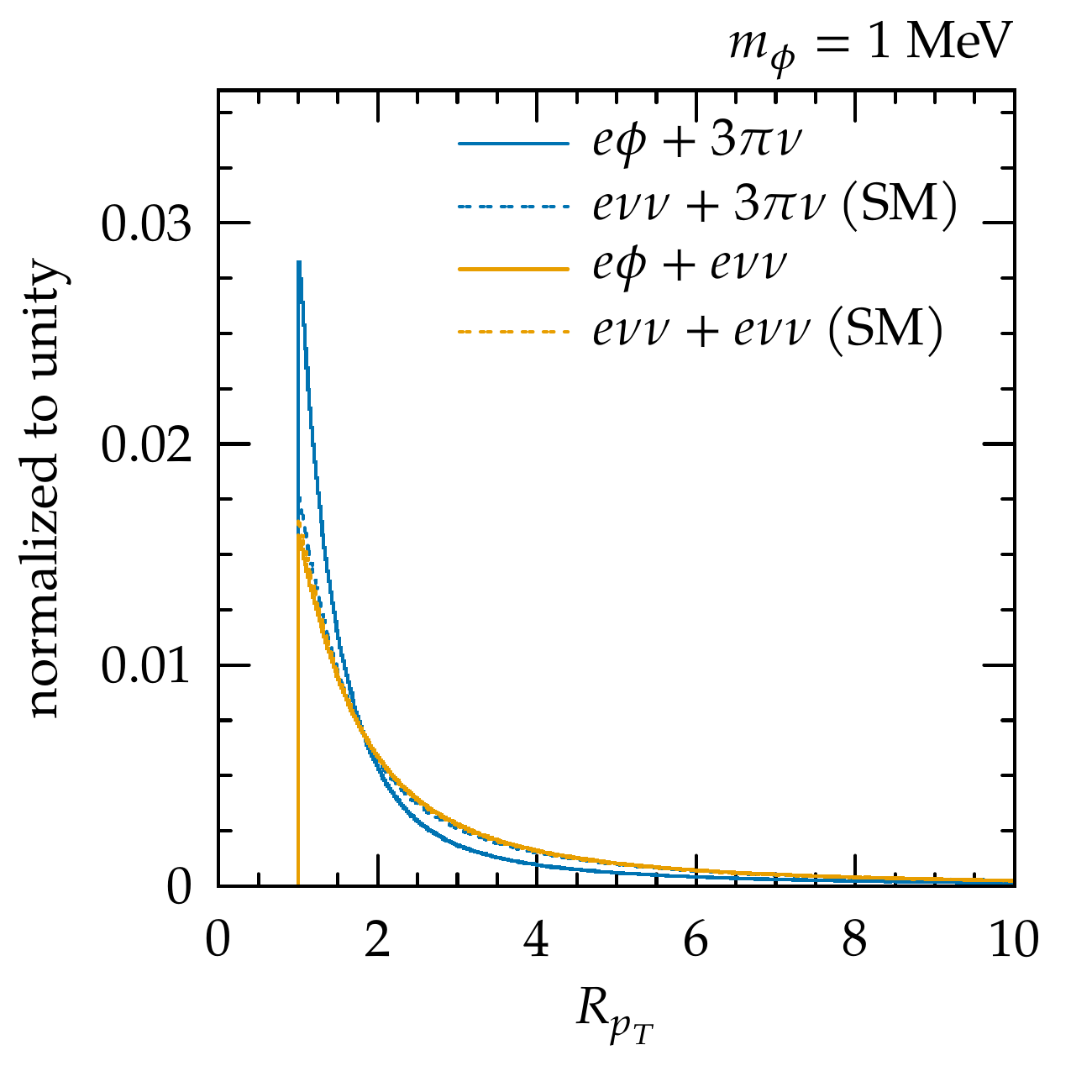}
  \end{center}
  \caption{\label{fig:xik}
Comparison between the ratio variables $\xi_{k,p}, R_{k_{\T}, p_{\T}}$ discussed below eq. (\ref{eq:xik}). Line conventions as in fig. \ref{fig:MT2_vs_M2}. We show the case $m_\phi = 1~\MeV$.
}
\end{figure}

Similarly as for $\MT2$, also the $\M2$ endpoint is the parent-particle mass. Compared to $\MT2$, $\M2$ distributions peak at a higher value and are more populated toward the endpoint. Interestingly, one common feature is that the smaller the number of invisible
particles, the more the distributions are populated toward their upper edge (for $\MT2$, see related discussions in \cite{Agashe:2010tu,Giudice:2011ib}).
This is displayed in fig. \ref{fig:MT2_vs_M2} for the case $m_\phi = 1~\MeV$.\footnote{\label{foot:figures_in_appendix}All of figures \ref{fig:MT2_vs_M2} through \ref{fig:xik} illustrate for a specific mass case, $m_\phi = 1~\MeV$, the variables discussed in the text that rely on the use of MAOS momenta. For the sake of comparison, analogous figures for $m_\phi = 1~\GeV$ are collected in fig. \ref{fig:variables_using_MAOS_1GeV} of the Appendix.} Hence a shape analysis of both $\MT2$ and $\M2$ could in principle be used to extract information about the number of invisible particles in the event, i.e whether the event is more signal- or background-like. Given the correlation between $\MT2$ and $\M2$, for definiteness we only consider the latter in the rest of our discussion.

Importantly, the MAOS solution to the constrained minimisation in eq. (\ref{eq:M2_ours}) can be used as an estimator of the {\em true} values of $\vb*{k}_{1,2}$, to be denoted as $\vb*{k}_{1,2}^{\maos}$ \cite{Cho:2008tj, Park:2011uz}. Similarly as in the $\MT2$ case, the $\M2$-based MAOS momenta \cite{Cho:2014naa,  Kim:2017awi} are distributed symmetrically around the true momenta, and peak at the respective true values.
A few remarks are in order. First, the $\M2$-based MAOS method turns out to be more efficient than the traditional MAOS method from $\MT2$~\cite{Kim:2017awi}. One reason is the fact that the $\MT2$-based MAOS solution comes with a twofold ambiguity in the longitudinal components of the invisible momenta, whereas the $\M2$-based MAOS solution is unique, as all momentum components are treated on a similar footing. Second, according to the definition of the full invariant masses $M$ in eq. (\ref{eq:M2_ours}), one further piece of information would be required: $k_{1,2}^2$. We set $k_{1,2}^2 = 0$, which usefully ensures the inequality $\M2 \le m_\tau$ for both signal and background events \cite{Cho:2008tj, Park:2011uz}. At face value, $k_{1,2}^2 = 0$ is a bad guess, because e.g. for the background channel $\ell \nu \nu$ one has $k^2 = m_{\nu \nu}^2$, which peaks around $1~\GeV^2$. We inspected more refined ansaetze, including the unfeasible case where one uses the truth-level $k_{1,2}^2$ values. We found that the $\M2$ distributions do not depend significantly on this choice, in agreement with existing literature \cite{Cho:2008tj, Park:2011uz}, although this issue may warrant further scrutiny.

With the thus-defined $\vb*{k}_{1,2}^{\maos}$ we can construct additional variables that would require knowledge of the invisible momenta, and that offer criteria for signal-background discrimination. A first example is $|\vb*{p}_\ell^{\tauRF}|$ in the $\oxt$ channel. In the $\tau$ rest frame (RF), obtained through either the MAOS or thrust methods, this quantity is equivalent to the signal-side invisible momentum. In fig. \ref{fig:pe} we thus compare the distribution for $p_e^{\maos} \equiv |\vb*{p}_e|^{\maos}_{{\tauRF}}$ with the corresponding quantity obtained through the thrust, for $\ma = 1~\MeV$, representative of the small-mass case.
For comparison, the case $\ma = 1~\GeV$ is shown in fig. \ref{fig:variables_using_MAOS_1GeV} in the Appendix.
In either case we restrict to the $\oxt$ channel.
MAOS and thrust achieve a comparable separation between the signal and the background distributions for $\ma = 1~\GeV$, whereas MAOS performs better for small $\ma$, as quantified by the full analysis, to be discussed at the end of the paper.

One further quantity constructible from $\vb*{k}_{1,2}^{\maos}$ is the ratio
\be
\label{eq:xik}
  \xik \equiv \frac{\min\{ |\k{1}|, \, |\k{2}| \}}{\max\{ |\k{1}|, \, |\k{2}| \}} \in [0,1]\,,
\ee
with $\xik^{\maos}$ denoting the corresponding ratio calculated with $\vb*{k}_{1,2}^{\maos}$.
This variable is reminiscent of the $\RpT$ ratio pointed out in Ref.~\cite{Agashe:2010tu}, with two differences. First, $\RpT$ is constructed in terms of the visible momenta, whereas $\xik$ requires the invisible ones, separately for the two branches -- which is precisely what MAOS provides.
Second, $\RpT$ is a `max-over-min' ratio, implying the non-compact domain $[1, \infty]$, i.e. a long distribution tail. The difference between signal and background is thus `diluted' over this tail. Conversely, $\xik$ spans the {\em compact} domain $[0,1]$, which enhances the shape difference between signal and background. The $\xik$ distribution performs best of all and is shown in fig.~\ref{fig:xik} (first panel) for the case $m_\phi = 1~\MeV$. Note that $\xi_{k,p}$ could be defined in the lab or CMS frames. We used the CMS-frame definition, where the slope differences are more pronounced.
For purposes of comparison, the $\xip$, $\RkT$ and $\RpT$ distributions are shown in the remaining panels of fig. \ref{fig:xik}. Besides, all of $\xi_{k,p}, R_{k_{\T},p_{\T}}$ are also shown in fig. \ref{fig:variables_using_MAOS_1GeV} of the Appendix for the case $m_\phi = 1~\GeV$. The underlying rationale of eq. (\ref{eq:xik}) is that this ratio is expected to be closer to unity for the $\oxo$-channel background (4th entry in the legend), because of the symmetric decay chains. By the same argument, we also expect the distribution `slope' to decrease roughly with the number of invisibles. The $\xik$ histograms, and to a lesser extent the $\xip$ ones, display both of these features.

\begin{figure}[t]
  \begin{center}
    \includegraphics[width=0.235\textwidth]{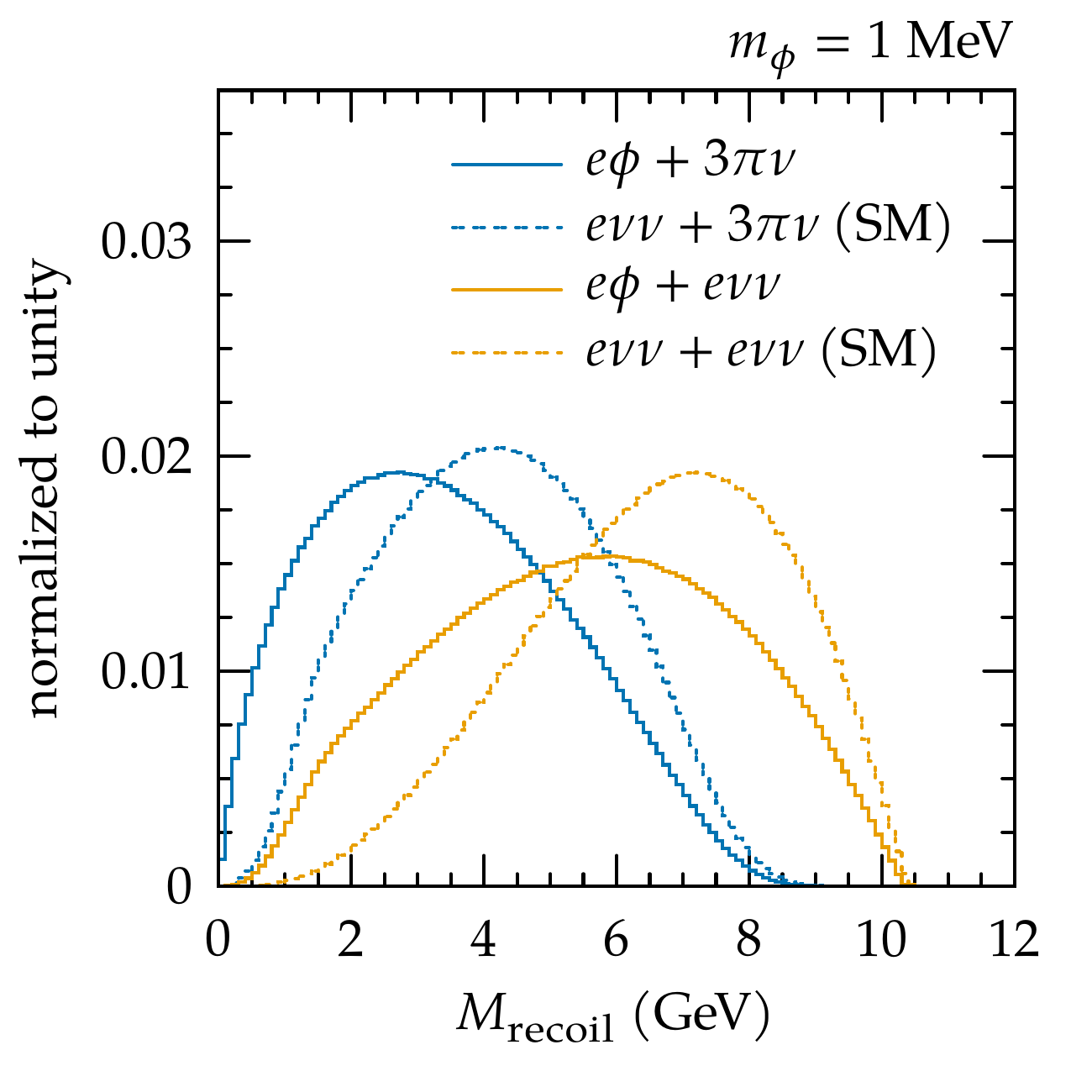}
    \includegraphics[width=0.24\textwidth]{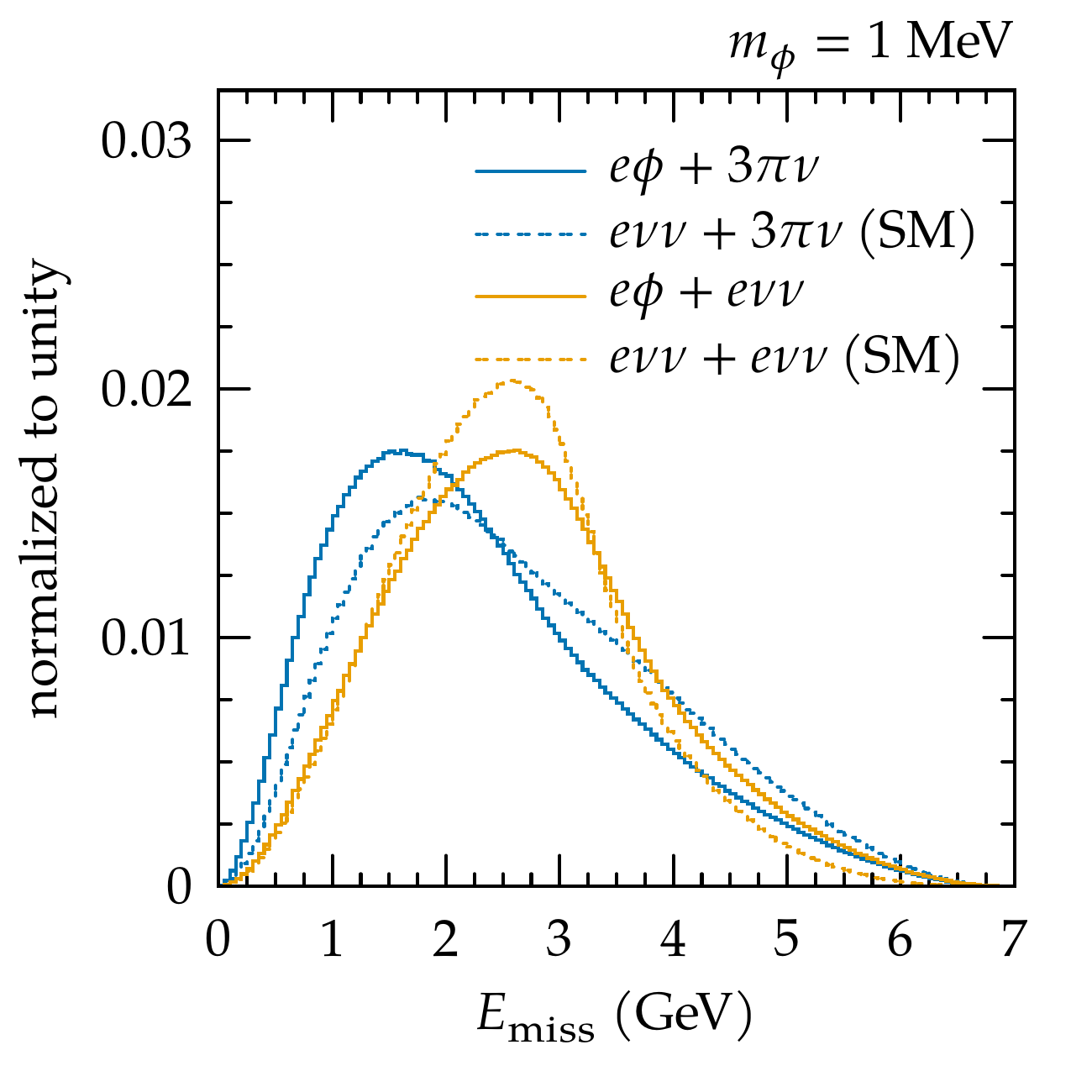}
  \end{center}
  \caption{\label{fig:Mrecoil_Emiss}
The variables $M_{\rm recoil}$ and $\Emiss$. Line conventions as in fig. \ref{fig:MT2_vs_M2}. We show the case $m_\phi = 1~\MeV$.
}
\end{figure}

We shortly discuss other known variables that do not require MAOS momenta, and that show a small enough correlation with those discussed so far. One popular example for lepton colliders is the recoil mass~\cite{Li:2012taa,Fujii:2015jha}, defined as $M_\text{recoil}^2 = ( P^\text{CMS} - p_1 - p_2 )^2$, i.e. the invariant mass of the full invisible system. Since there are more invisible particles in the  background, they `typically' have a larger invariant mass than in the signal case. This property is clearly visible in the $M_{\rm recoil}$ distribution in fig. \ref{fig:Mrecoil_Emiss} (first panel), for both the $\oxt$ and the $\oxo$ cases.
One may reverse the argument for the variable $\Emiss \equiv |\vb*{P}_\text{miss}|$: in the $\tau$-pair rest frame, the invisible particles would be boosted along the momentum direction of the parent $\tau$, so their three-momenta will partially cancel those from the other decay chain. This cancellation will be the more efficient, the more symmetric is the decay. Hence one may expect a `thicker' tail for the signal decay $\ell \phi + \ell \nu \nu$, than for the corresponding background $\ell \nu \nu + \ell \nu \nu$. In practice, the $\Emiss$ discriminating power is inferior to $M_{\rm recoil}$'s, as shown by the second panel of fig.~\ref{fig:Mrecoil_Emiss}.\footnote{
Other examples of variables relying on the visible kinematics only were investigated in \cite{DeLaCruz-Burelo:2020ozf}. We do not include them in our analysis, as we do not expect them to modify appreciably our conclusions, as is also the case for $\Mrecoil$ and $\Emiss$.
}

Before presenting our main analysis, we collect details about our setup. We generate $e^+ e^- \to \tau^+ \tau^-$ using {\tt MadGraph}. Tag-side decays and backgrounds are obtained through {\tt TauDecay} \cite{Hagiwara:2012vz}, whereas signal-side decays are populated as phase space through {\tt ROOT}. We generate about $1.5\cdot 10^7$ events per process. In order to populate the phase space in a similar way as Ref. \cite{Tenchini:2020njf}, we apply the cut $0.8 \le T \le 0.99$ on the thrust scalar. We also inspected the effect of including further cuts on the total visible energy and on the invariant mass of the 3-prong system, used in the Belle-II analysis \cite{Tenchini:2020njf} for the suppression of reducible backgrounds, and found it to be negligible in our case. Momentum smearing due to detector effects is typically $\lesssim 1$\% and we safely neglect this effect. To be more exact, event distributions are vastly more populated for $p_T > 0.2$ GeV/c \cite{BelleIITrackingGroup:2020hpx}, which is the region where the below-1\% momentum-smearing figure holds. Our numerical analysis uses the public library {\tt YAM2}~\cite{Park:2020bsu} and the {\tt TMVA} \cite{Hocker:2007ht} class available in {\tt ROOT}. We restrict to phase-space decays for comparison with Ref. \cite{Tenchini:2020njf}, and also for the following reason.
The variables discussed above are insensitive to the angular distribution of the new particle.
The cuts implemented to mimic the search in Ref.~\cite{Tenchini:2020njf} will not modify angular distributions either, as these cuts affect invariant masses or momentum magnitudes. We also note that, since our decay of interest is $2$-body, the decay amplitude is isotropic and so is the differential decay rate \cite{Zyla:2020zbs}. As a consequence, our results apply equally to the case of an ALP and of a hidden photon, with coupling chirality whatever. A separate, interesting question would be to construct ``invisible-spin-savvy'' variables, which exploit MAOS or thrust momenta in spin-sensitive kinematic variables (as in e.g. \cite{Guadagnoli:2013xia}), in order to tell apart different coupling assumptions.

\begin{figure*}[t]
  \begin{center}
    \includegraphics[width=0.245\textwidth]{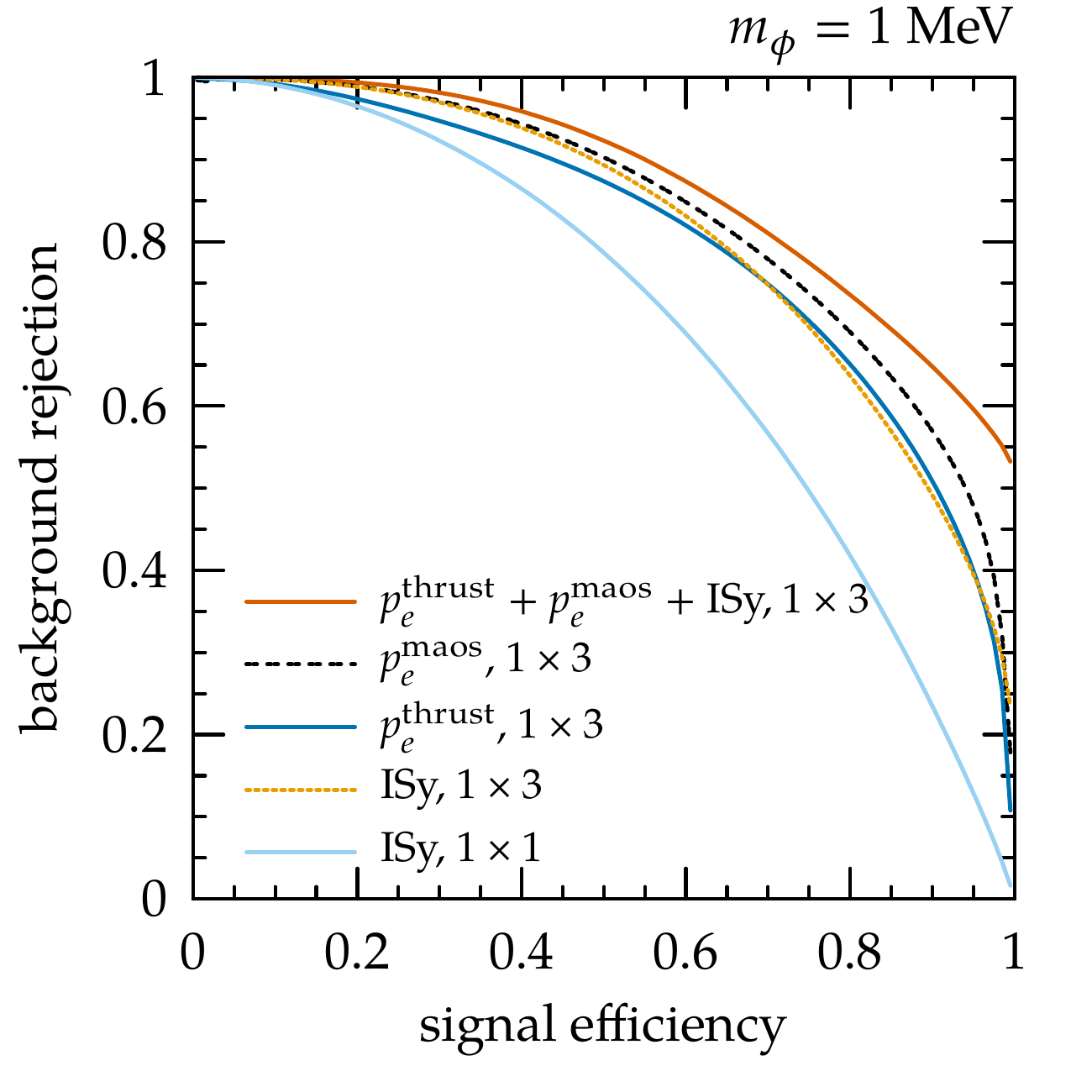}
    \includegraphics[width=0.475\textwidth]{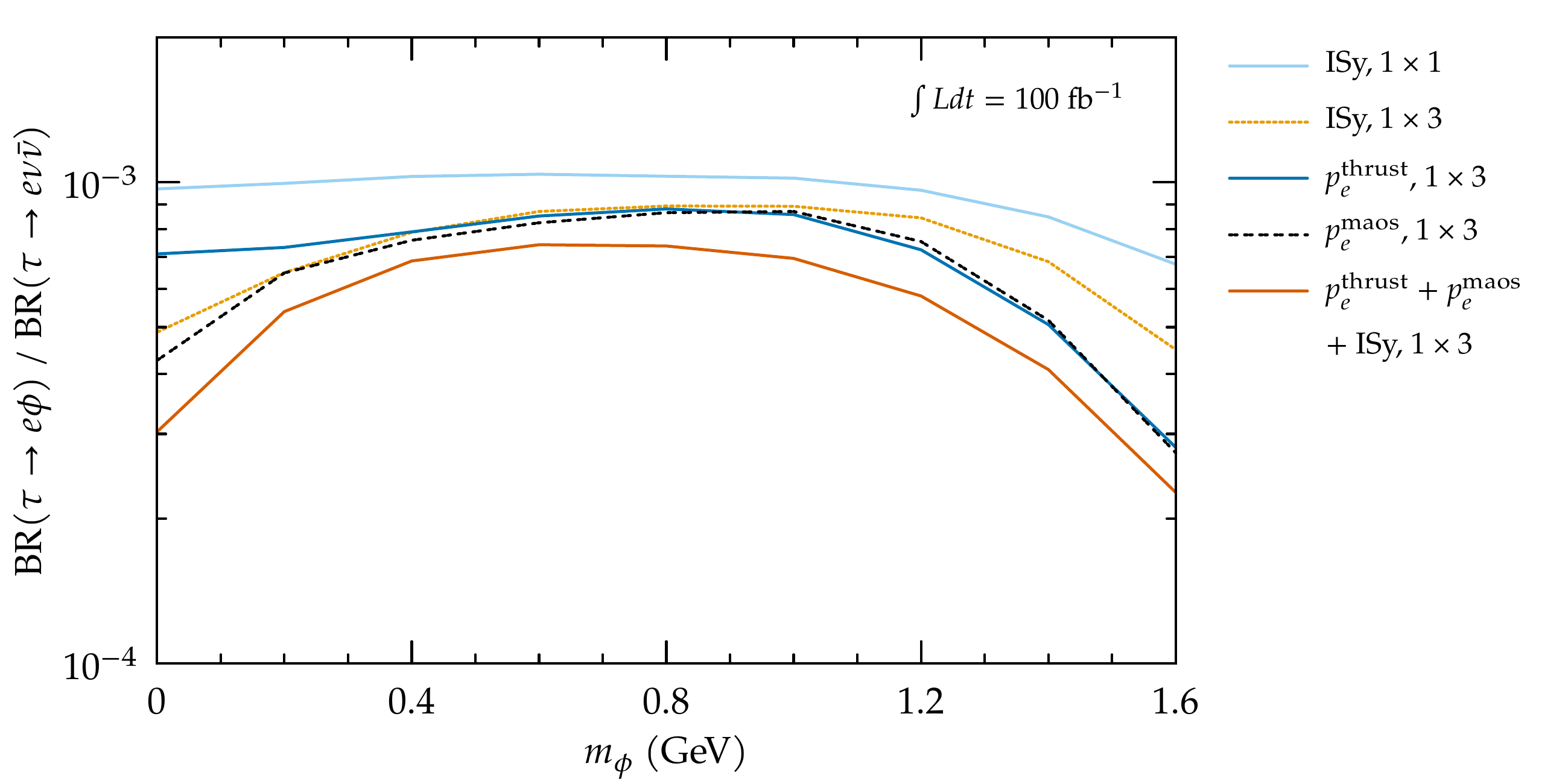}
    \includegraphics[width=0.245\textwidth]{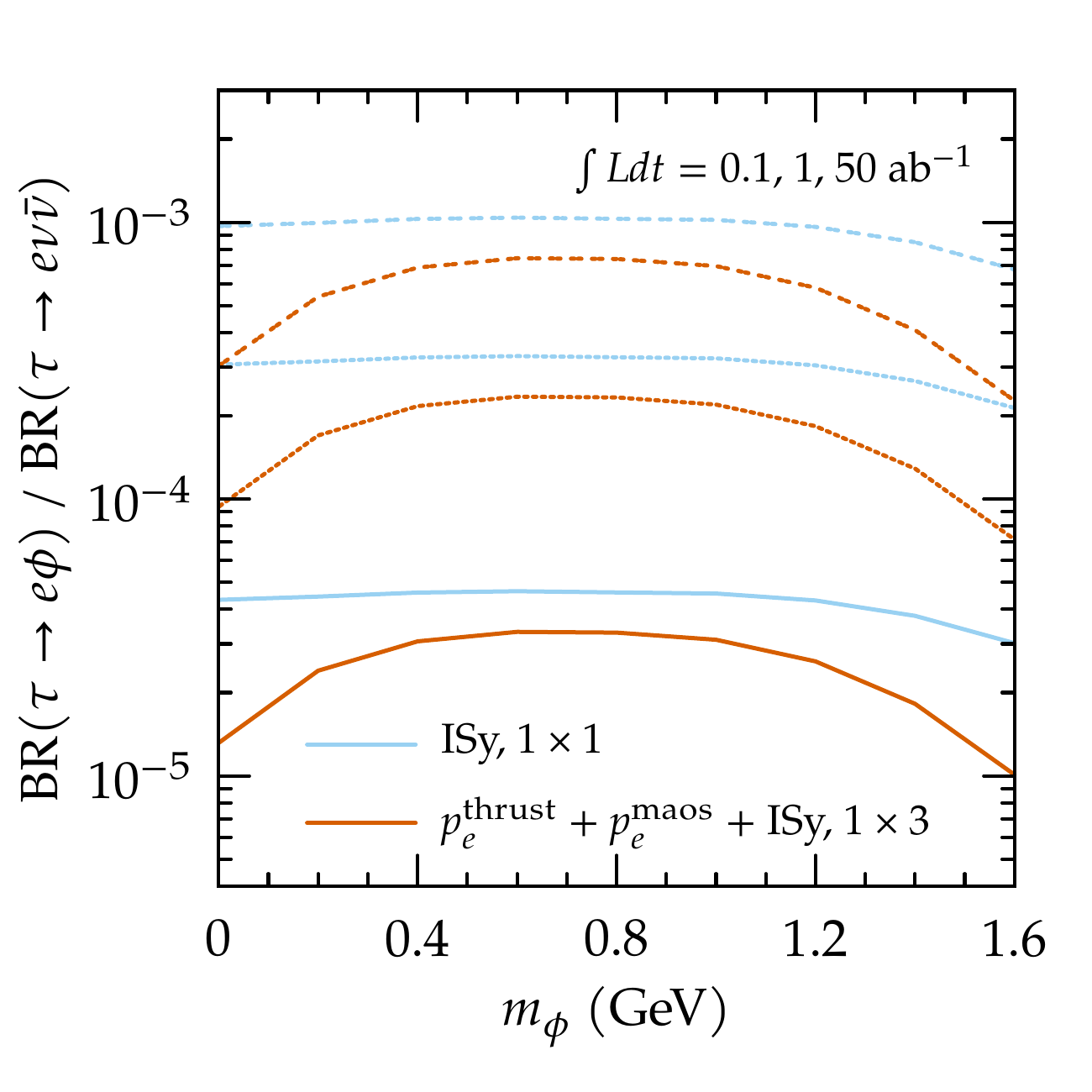}
  \end{center}
  \caption{Left: ROC curves for the different classifiers discussed in the text, in the case $\ma = 1~\MeV$. Right: 95\% CL upper limits on the signal branching ratios in units of the $\tau \to \ell \nu \nu$ one. We consider the different classifiers discussed in the text (see legend) and for each of them three reference luminosities (see plot headers).
\label{fig:ROC_UL}
}
\end{figure*}

\newcommand{\ourC}{{\tt ISy}}
We next discuss the main analysis. We first note that $\M2$, $\xi_{k,p}$, $\Mrecoil$ and $\Emiss$ can be unambiguously calculated for the $\oxt$ and $\oxo$ cases alike,\footnote{\label{foot:1x1}On the other hand, application of $p_e^{\maos}$ and $p_e^{\rm thrust}$ is not straightforward in the $\oxo$ case, because the symmetry of the topology implies a combinatorial ambiguity, introducing a separate source of uncertainty. Including these variables in this case requires a dedicated study (see e.g. discussion in \cite{Choi:2011ys}).} and their distributions depend, to different degrees, on the number of invisibles in the decay. We thus collectively denote this ensemble as `invisible-savvy' variables, and construct a classifier that we refer to as $\ourC$.
\newcommand{\thrust}{{\rm thrust}}
\newcommand{\pem}{p_e^\maos}
\newcommand{\pet}{p_e^\thrust}
Note that $\ourC$ does {\em not} include $\pem$.
We then consider the following cases: 
{\em (a)} $\pet$ alone, on $\oxt$ decays; 
{\em (b)} $\pem$ alone, on $\oxt$ decays; 
{\em (c)} $\ourC$ alone, on $\oxt$ decays; 
{\em (abc)} $\pet$ + $\pem$ + $\ourC$ combined, on $\oxt$; 
{\em (d)} $\ourC$, on $\oxo$.
With case {\em (a)} we reproduce the results in Ref. \cite{Tenchini:2020njf} and thus validate our setup; the comparisons {\em (a)} vs. {\em (b)} vs. {\em (c)} vs. {\em (abc)} show the improvement achievable on the $\oxt$ channel alone with the different variables, and with their combination; {\em (c)} vs. {\em (d)} shows the relative performance of the two considered channels. In this last comparison, we only consider the $\ourC$ classifier, keeping in mind footnote \ref{foot:1x1}.

We refrain from including an {\em (abcd)} case, which would show the `maximal' improvement achievable from both the use of two channels in lieu of one, and the use of more variables. In fact, a reliable combination of $\oxt$ and $\oxo$ is not straightforward, because of the different tag, and should be performed on actual data.

Fig. \ref{fig:ROC_UL} (first panel) presents a comparison of the performance profiles for the cases discussed above, i.e. {\em (a)} to {\em (d)}, in the plane of signal efficiency vs. background rejection, yielding the `ROC' curve. In the single-variable cases {\em (a)} and {\em (b)}, we use a cut-based approach rather than a BDT/NN, although the optimal cut is obtained by {\tt TMVA}. We see that, for $\ma = 1$ MeV, $\pem$ has a larger area-under-curve (AUC) than $\pet$ for any signal efficiency. For $\ma = 1$ GeV, the two AUCs are comparable, and we show the corresponding plot in fig. \ref{fig:BDT_ROC1GeV} (left) in the Appendix. In the remaining panels of this figure we also show, for the sake of possible reproducibility, the BDT response for both cases $\ma = 1~\MeV$ and $1~\GeV$.

Our $S/B$ separation can be translated into an estimate of the upper limit (UL) on $\mc B(\tau \to e \phi)$ achievable with a given Belle-II luminosity. To determine such limit we proceed as follows. Given a sample of $N_s$ signal and $N_b$ background events, the weight of the sample background can be determined as $w_b = B / N_b$ where, in our case, $B = \sigma_{\tau \tau} \times \mc B(\tau\tau \to {\rm bkg}) \times \mc L$, with $\sigma_{\tau \tau}$ the total $\tau^+ \tau^-$ production cross-section and $\mc L$ the luminosity.
We need the weight $w_s$ for the signal sample, from which we may estimate the statistical significance as $\sigma(w_s) \simeq S / \sqrt{S + B} = w_s N_s / \sqrt{w_s N_s + w_b N_b}$. The $\sigma(w_s)$ value corresponding to a 95\% confidence-level (CL) exclusion is given by $\sigma(w_s) = 1.96$ \cite{Zyla:2020zbs}, that we invert in terms of $w_s$. (With fixed $N_{s,b}$, one may proceed iteratively starting from $w_s^{(0)} = w_b$ and decreasing $w_s^{(i)}$ till the desired equality is satisfied.) 
From $S = w_s N_s$, one finally obtains the corresponding signal branching-ratio value through $\mc B(\tau \to e \phi) / \mc B(\tau \to {\rm bkg}) = S / B$.

This procedure is independent from the template-fit method used for the on-going Belle-II analysis \cite{Tenchini:2020njf}, hence the agreement with \cite{Tenchini:2020njf} of the $\pet$-case upper-limit curve shown in fig.~\ref{fig:ROC_UL} is a non-trivial check of our approach. We also verified that, after the classifier cut, $B$ is large enough that the above approximate relations are valid~\cite{Bhattiprolu:2020mwi}.

Before concluding with the main analysis results we note that our approach has a wide range of applicability -- be it to searches of new decays to invisibles or to improving the knowledge of background decays to invisibles -- to the extent that the `pairwise-decay' topology is the same.  
Besides, if one restricts to transverse variables, a similar approach may also be applied for meson or $\tau$ decays at hadron colliders.

We conclude by presenting the 95\% CL upper limit on the signal branching ratio (BR) as a function of the new light particle mass, for different assumed Belle-II luminosities $\mc L$, and with the different classifiers discussed. These upper limits are summarised in the middle and rightmost panels of fig. \ref{fig:ROC_UL}. In particular, the former shows, for the integrated luminosity $\mc L = 0.1$ ab$^{-1}$, a comparison among the cases {\em (a)}, {\em (b)}, {\em (c)}, {\em (d)}, {\em (abc)} discussed above; conversely, the last panel of fig. \ref{fig:ROC_UL} focuses on the evolution of the expected upper limit with luminosity, and shows the cases $\mc L = \{ 0.1, 1, 50 \}~\mbox{ab}^{-1}$, corresponding to the dataset accumulated as of Summer 2021, the dataset anticipated before the 2022 shutdown, and the target Belle-II dataset, respectively. As also suggested by the ROC curves, $\pem$ allows for a better BR limit than $\pet$ for small $\ma \lesssim 0.1$ GeV. For example, this improvement is a factor of $\approx 1.8$ for $\ma \approx 0$ and $\mc L = 0.1/$ab.
In short, for an ALP or hidden vector of small $\ma \lesssim 1$ MeV, we anticipate that application of our full strategy to the $\oxt$ channel alone will lead to a 95\%-CL limit of around
\bea
\mc B(\tau \to e \phi) &\le& \{ 5.4 \cdot 10^{-5}, 1.7 \cdot 10^{-5}, 2.4 \cdot 10^{-6} \},\nn \\
\mbox{for}~~~ \mc L &=& \{ 0.1, 1, 50 \}~\mbox{ab}^{-1},
\eea
to be compared with $\mc B(\tau \to e \phi) \le \{ 
1.3 \cdot 10^{-4}, 4.0 \cdot 10^{-5}, 5.7 \cdot 10^{-6}
\}$ with the thrust method alone. As a consequence, our strategy {\em improves by a factor close to 3} the limit achievable with the strategy currently in place within Belle II. We thus expect a Belle-II limit on BR$(\tau \to e$ + invisible$)$ stronger than the existing ARGUS limit \cite{Albrecht:1995ht} by a factor of respectively 50, 170, 1150 with 0.1, 1, 50 ab$^{-1}$ Belle-II data.

\acknowledgments

\noi DG warmly acknowledges Diego Redigolo for several related discussions following his talk \cite{Redigolo:GDR}, and Justine Serrano for comments. Exchanges with Aleks Smolkovic and Jure Zupan are also acknowledged.
This project has received funding from the ANR under contract n. 202650 (PRC `GammaRare'), IBS under the project code, IBS-R018-D1 and from the European Union’s Horizon 2020 research and innovation programme under the Marie Sk{\l}odowska-Curie grant agreement No 101026516.

\appendix

\section*{Additional Figures}

\begin{figure*}[t]
  \begin{center}
    \includegraphics[width=0.24\textwidth]{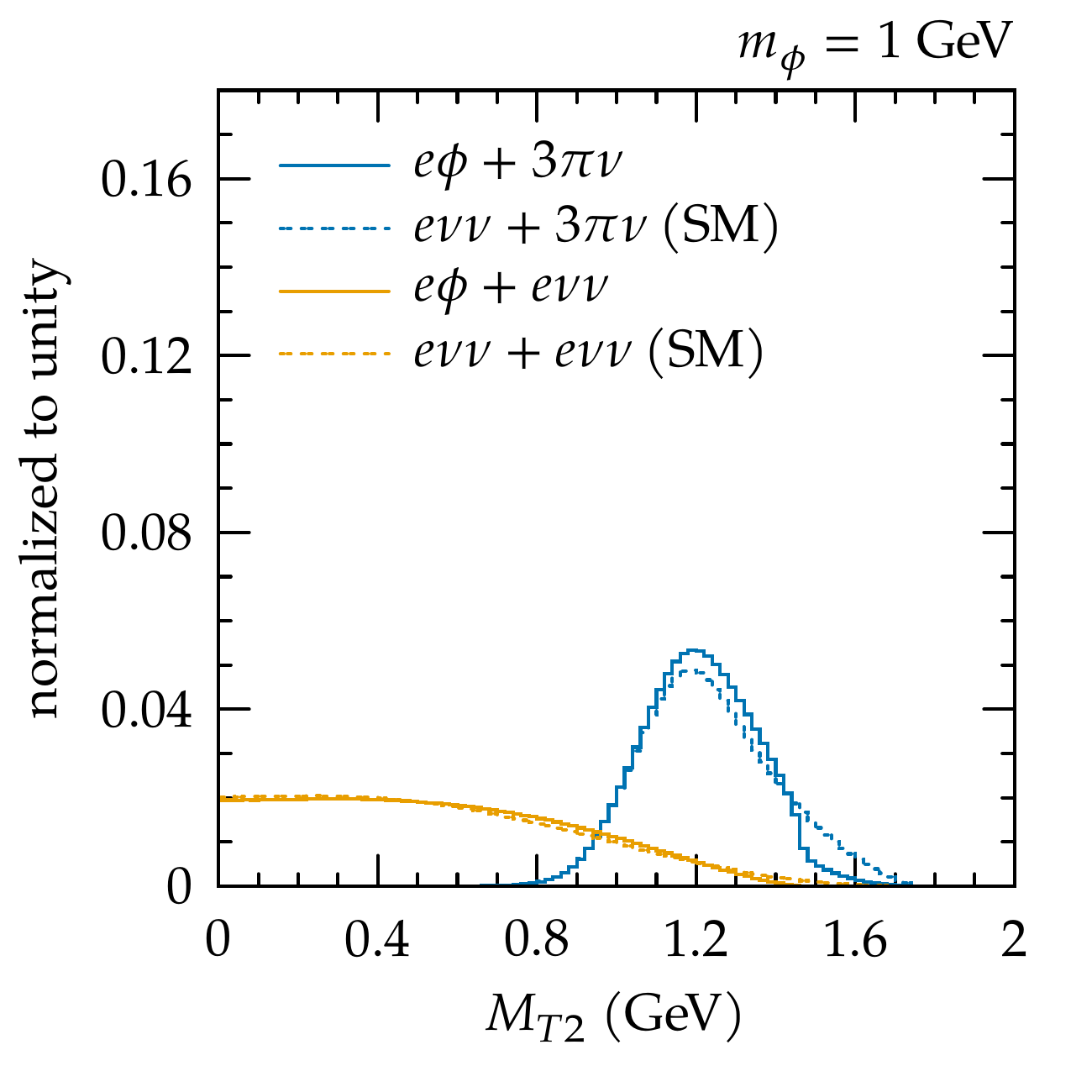}
    \includegraphics[width=0.24\textwidth]{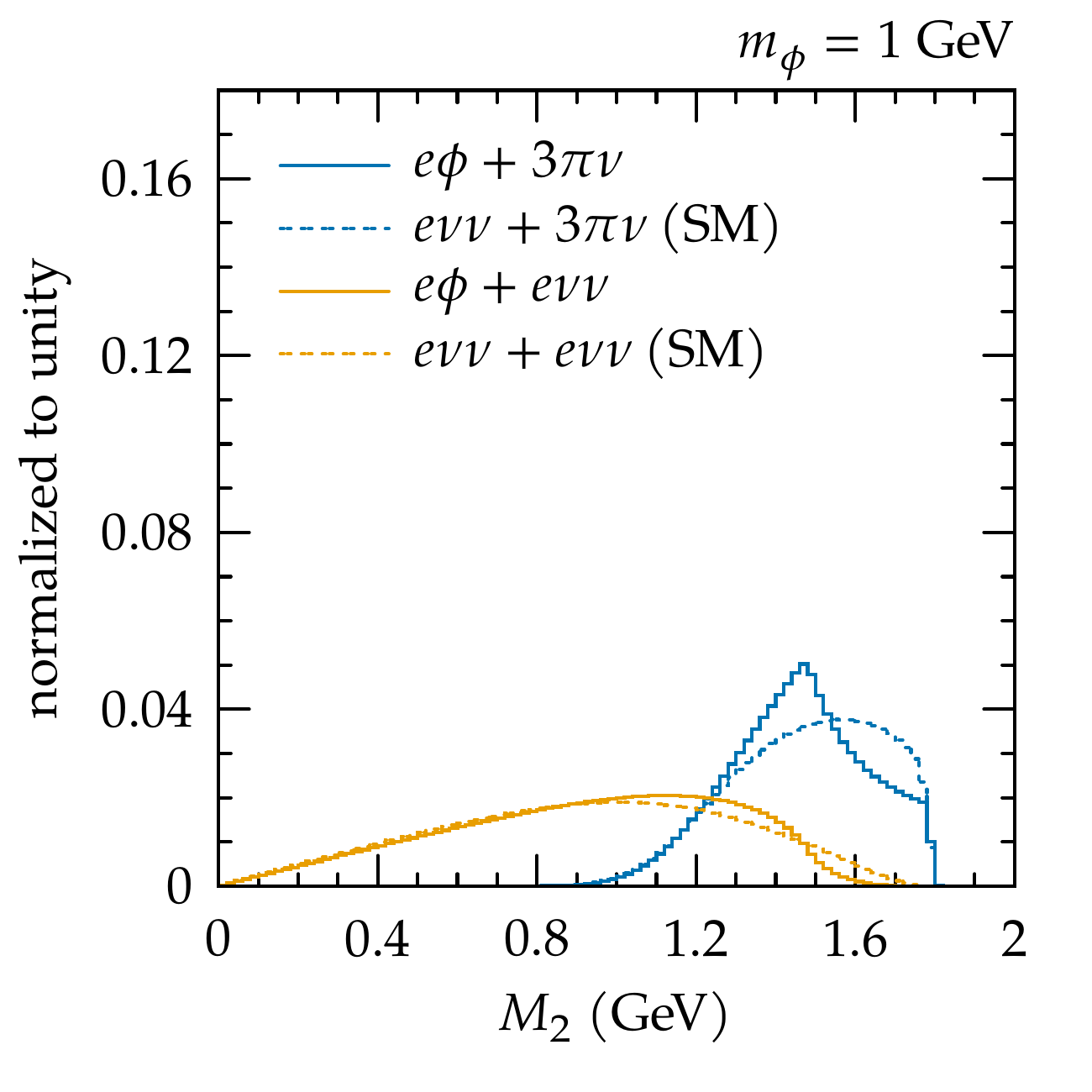}
    \includegraphics[width=0.24\textwidth]{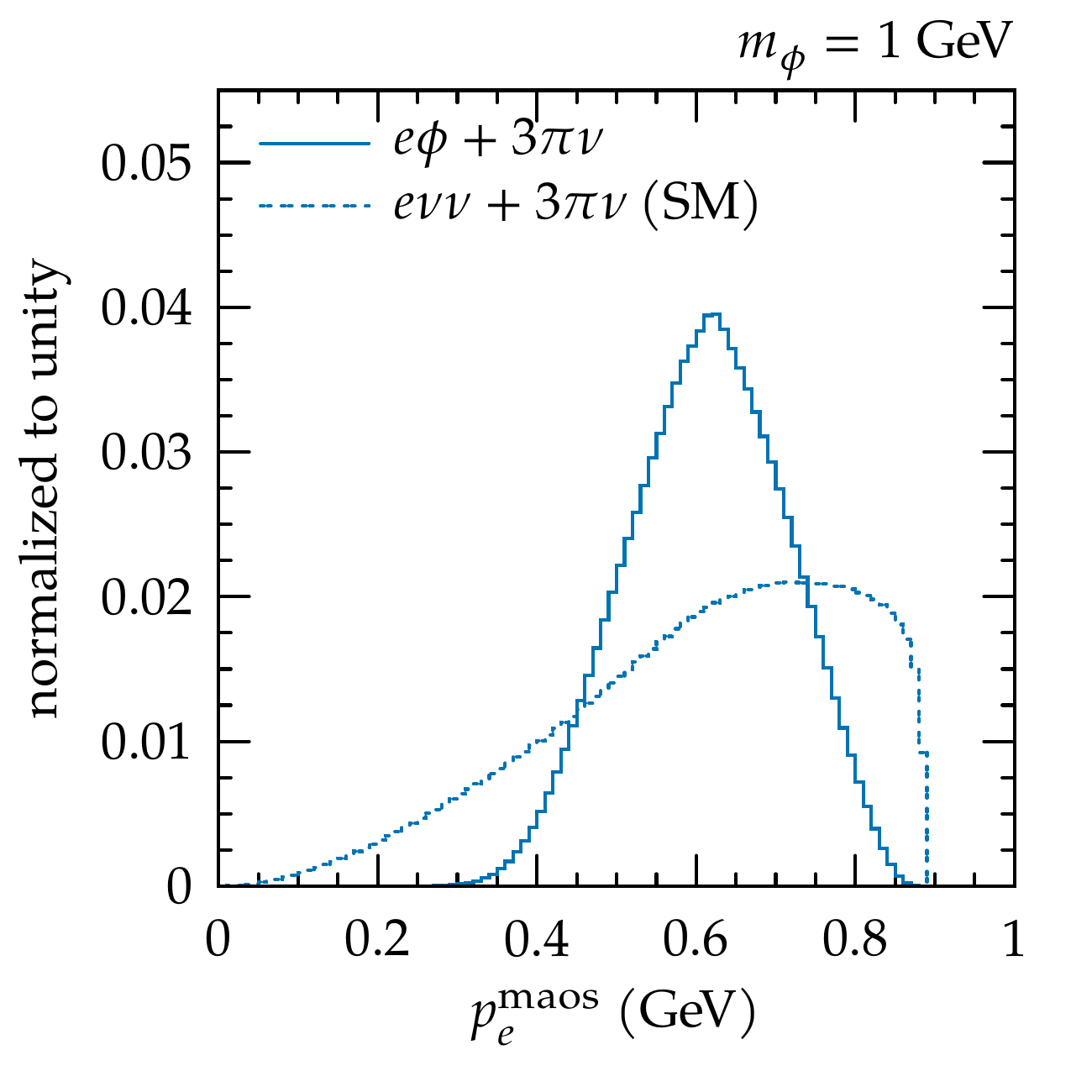}
    \includegraphics[width=0.24\textwidth]{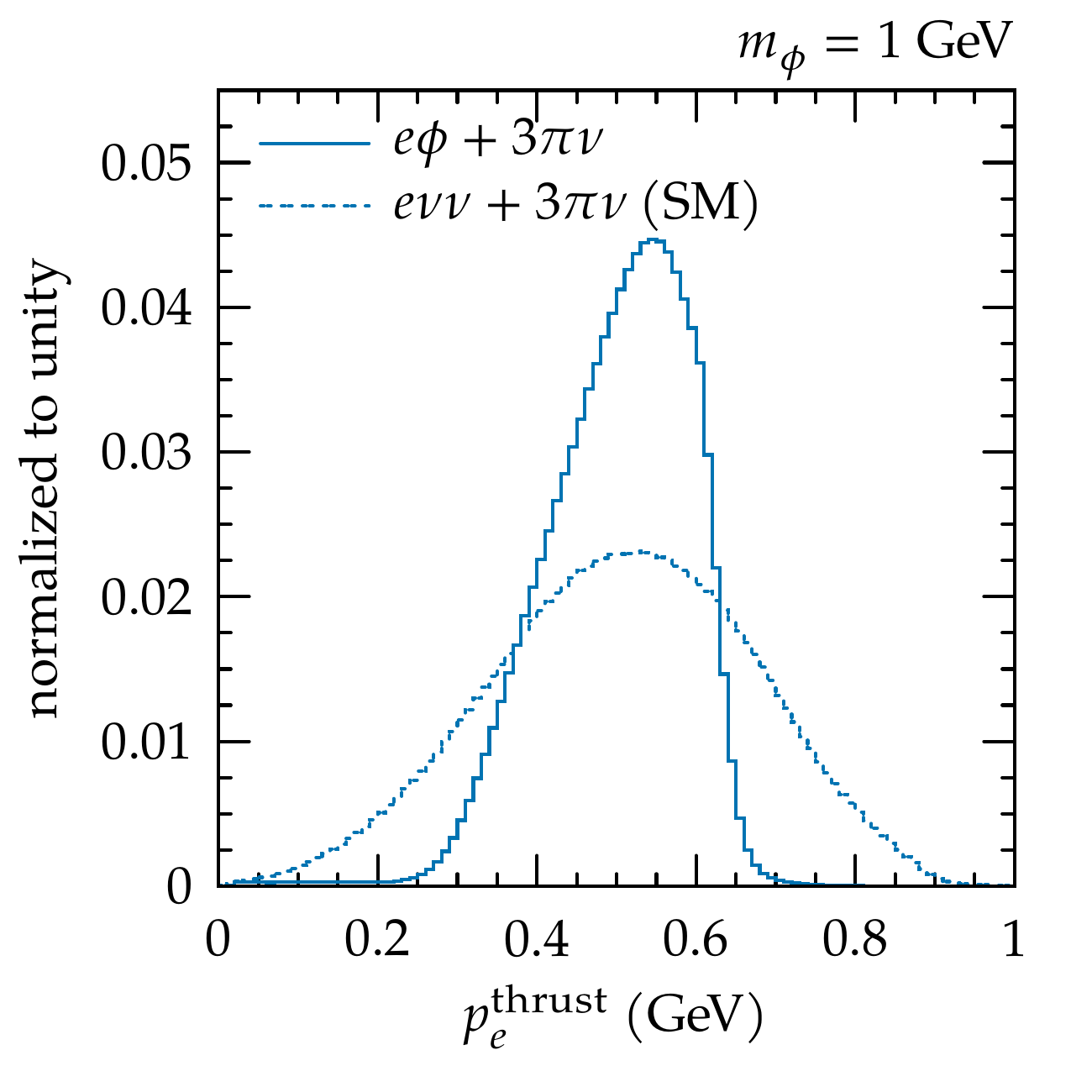}
    \includegraphics[width=0.24\textwidth]{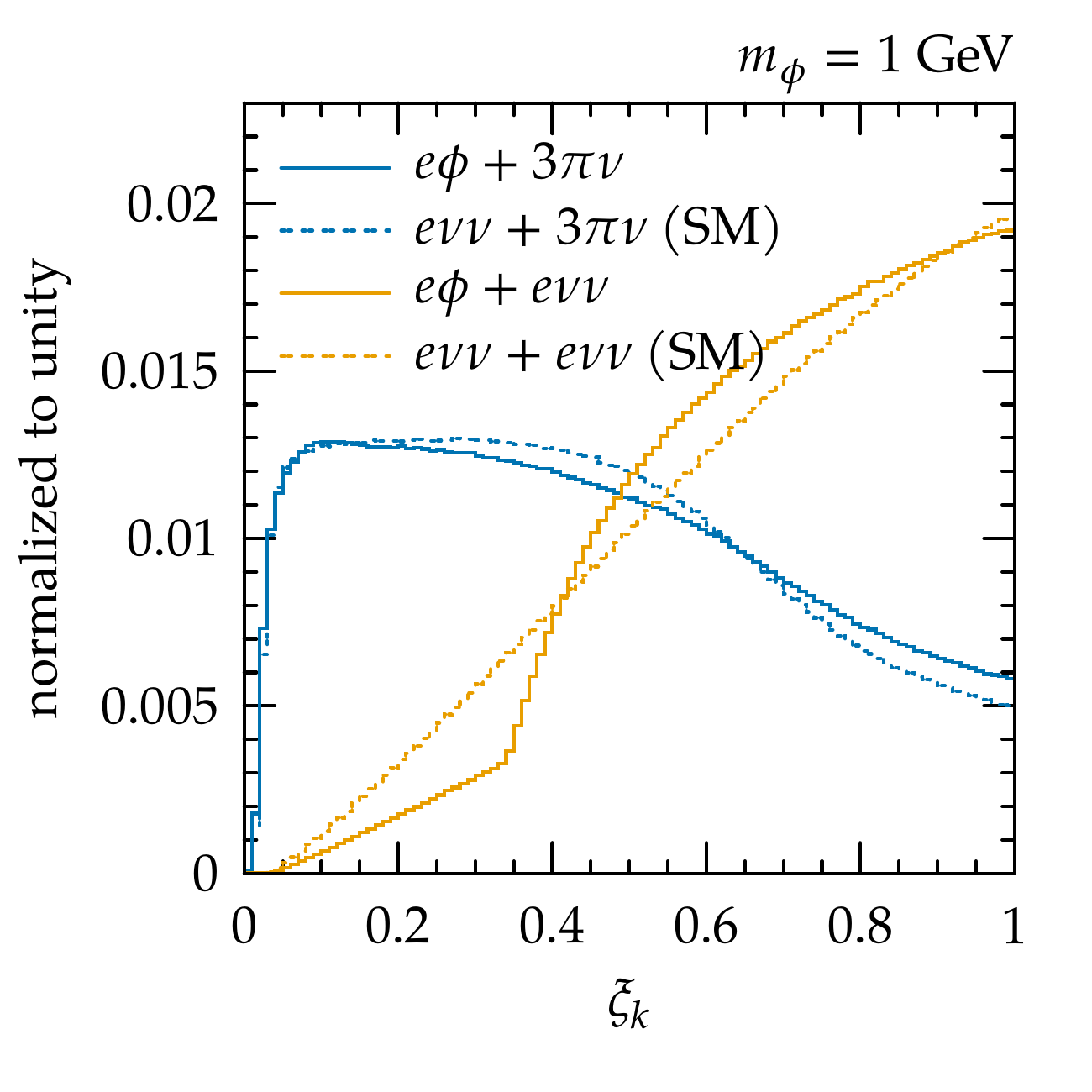}
    \includegraphics[width=0.24\textwidth]{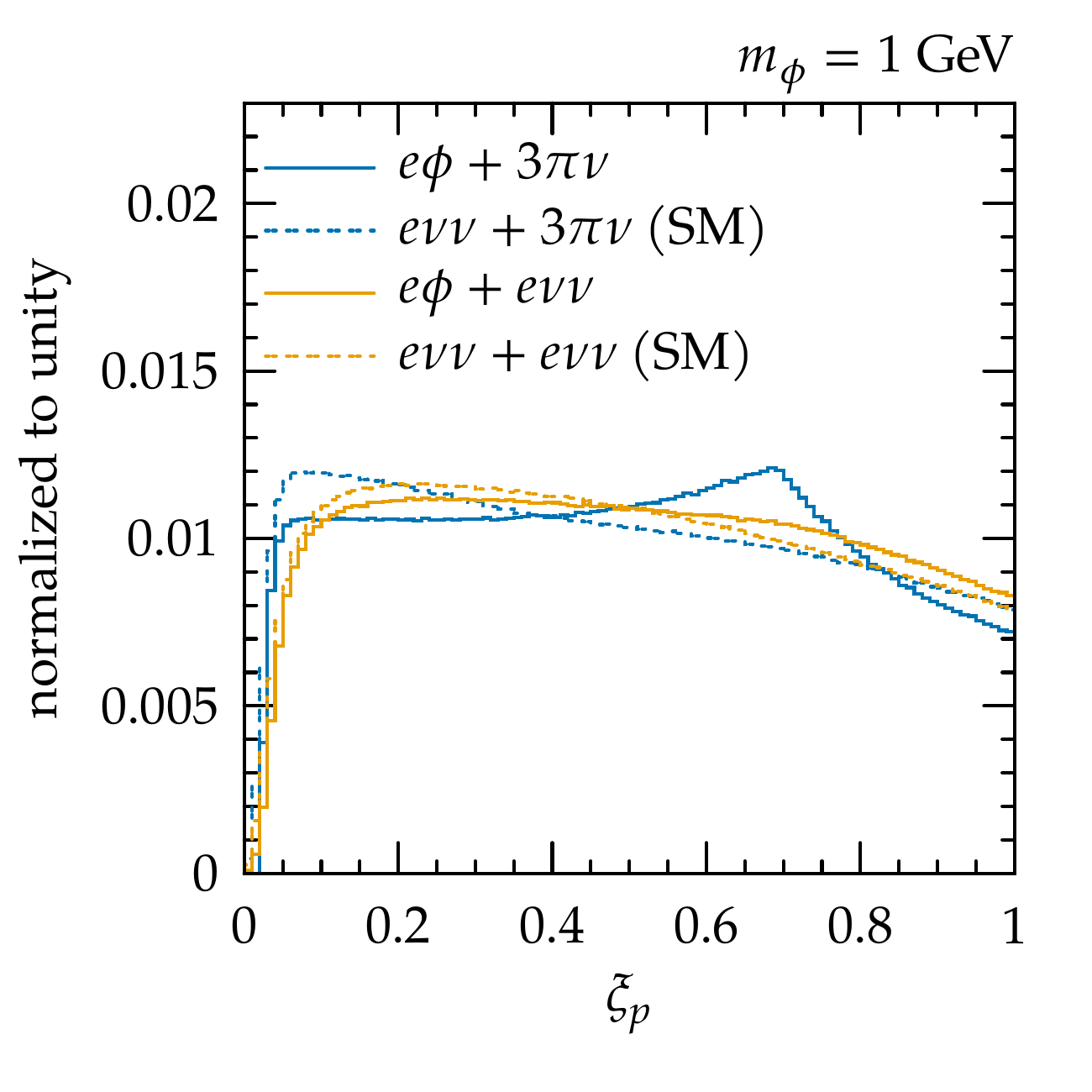}
    \includegraphics[width=0.24\textwidth]{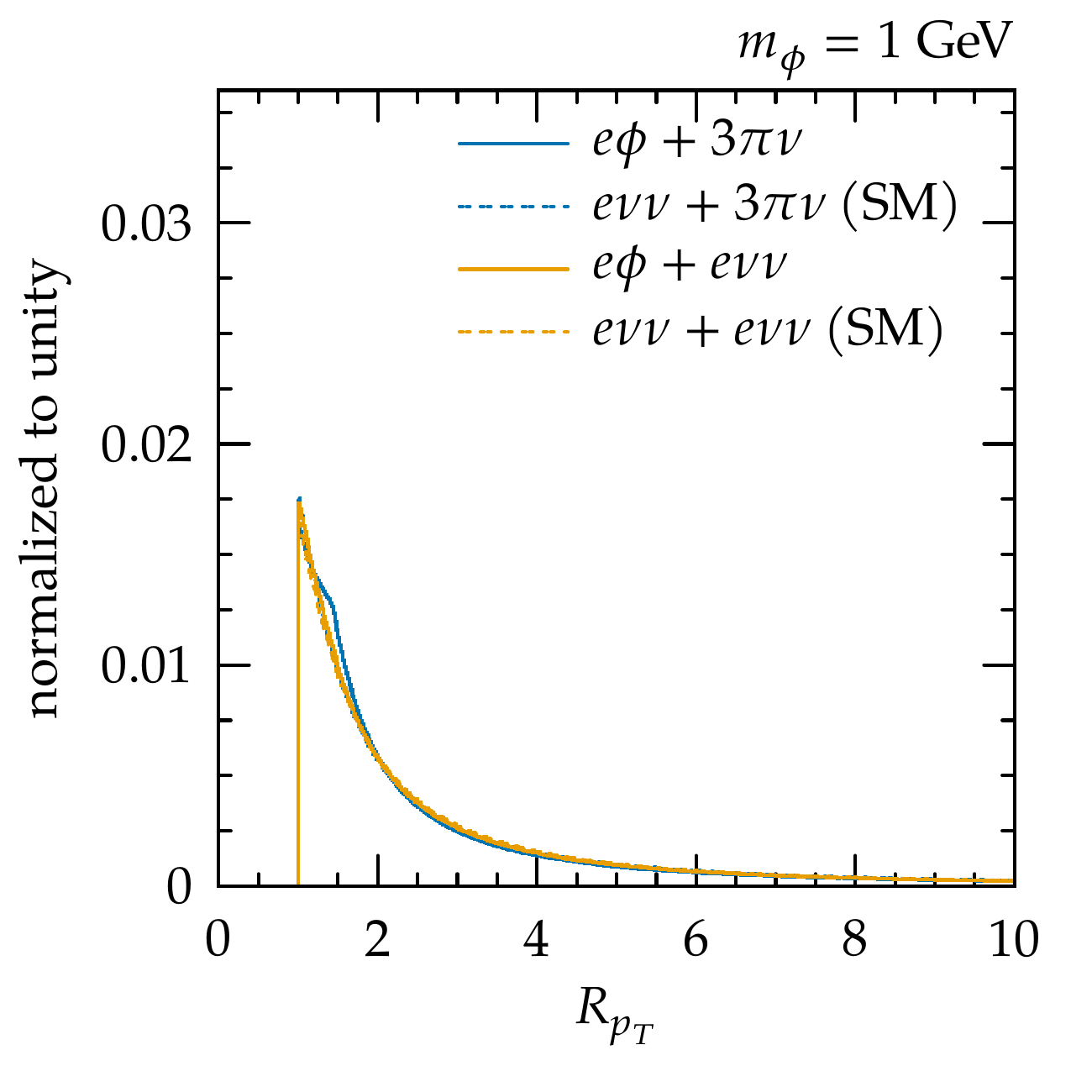}
    \includegraphics[width=0.24\textwidth]{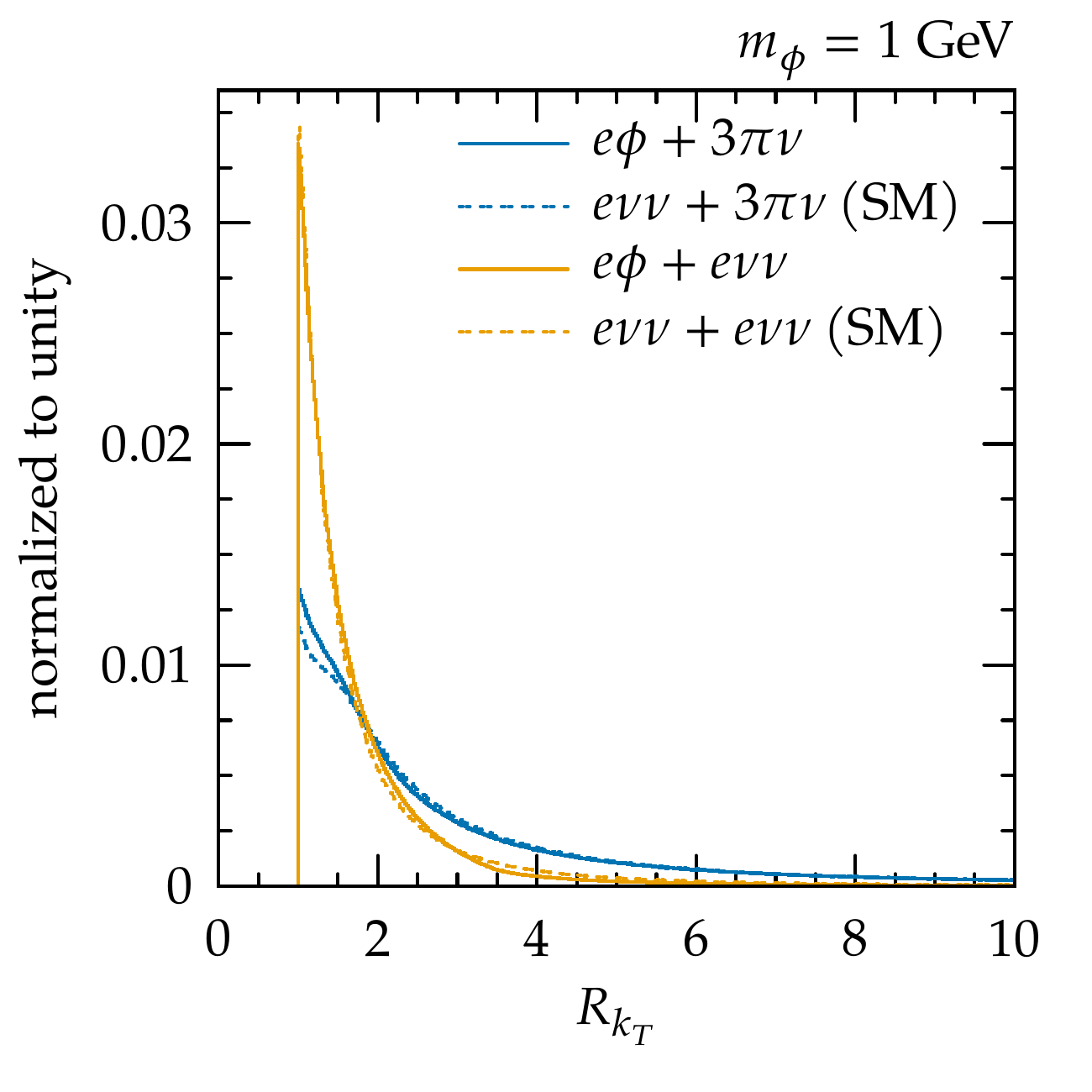}
  \end{center}
  \caption{\label{fig:variables_using_MAOS_1GeV}
Same as figs. \ref{fig:MT2_vs_M2} + \ref{fig:pe} + \ref{fig:xik} (with the corresponding panels in the same order), but for $\ma = 1$ GeV.}
\end{figure*}

\begin{figure*}[t]
  \begin{center}
    \includegraphics[width=0.30\textwidth]{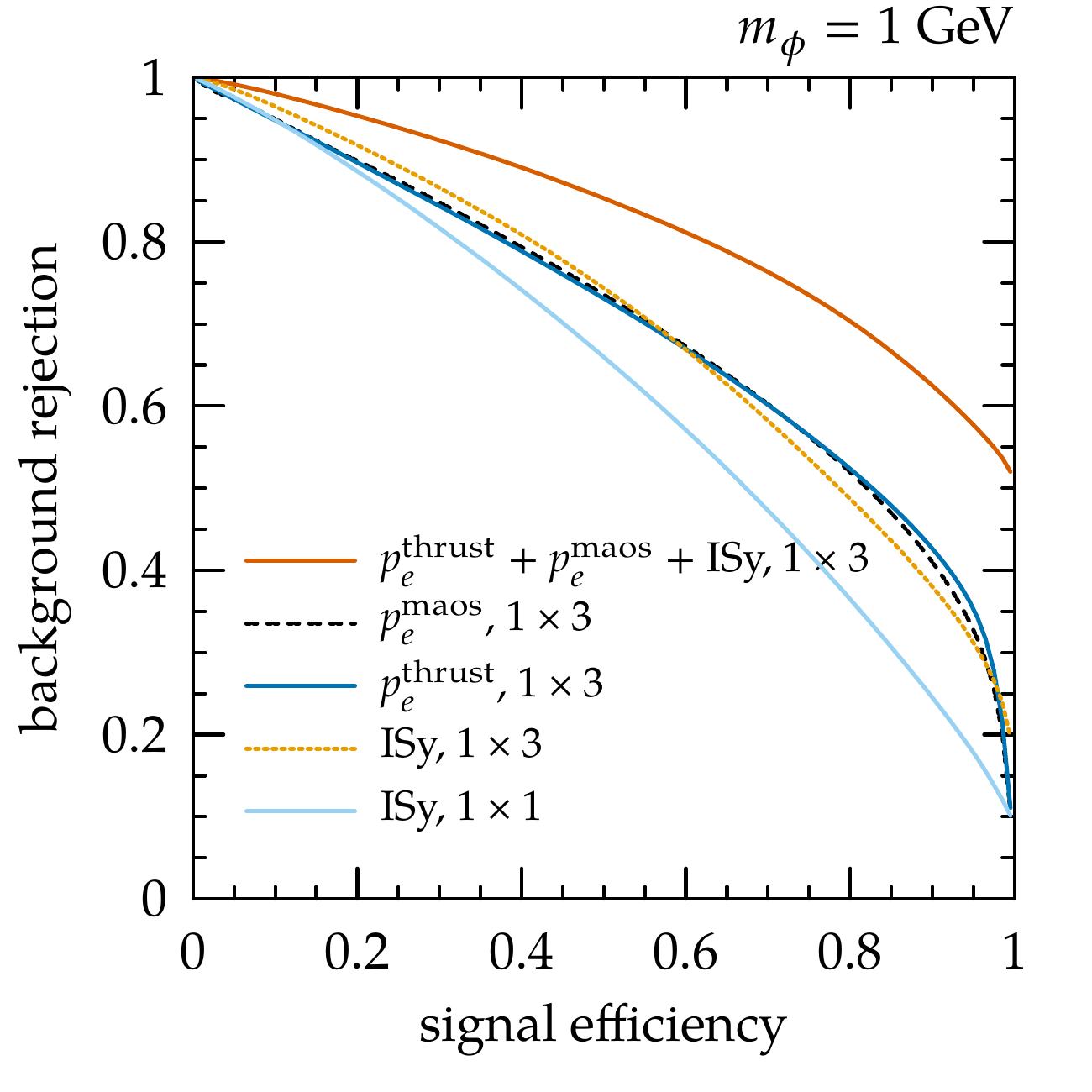}
    \includegraphics[width=0.30\textwidth]{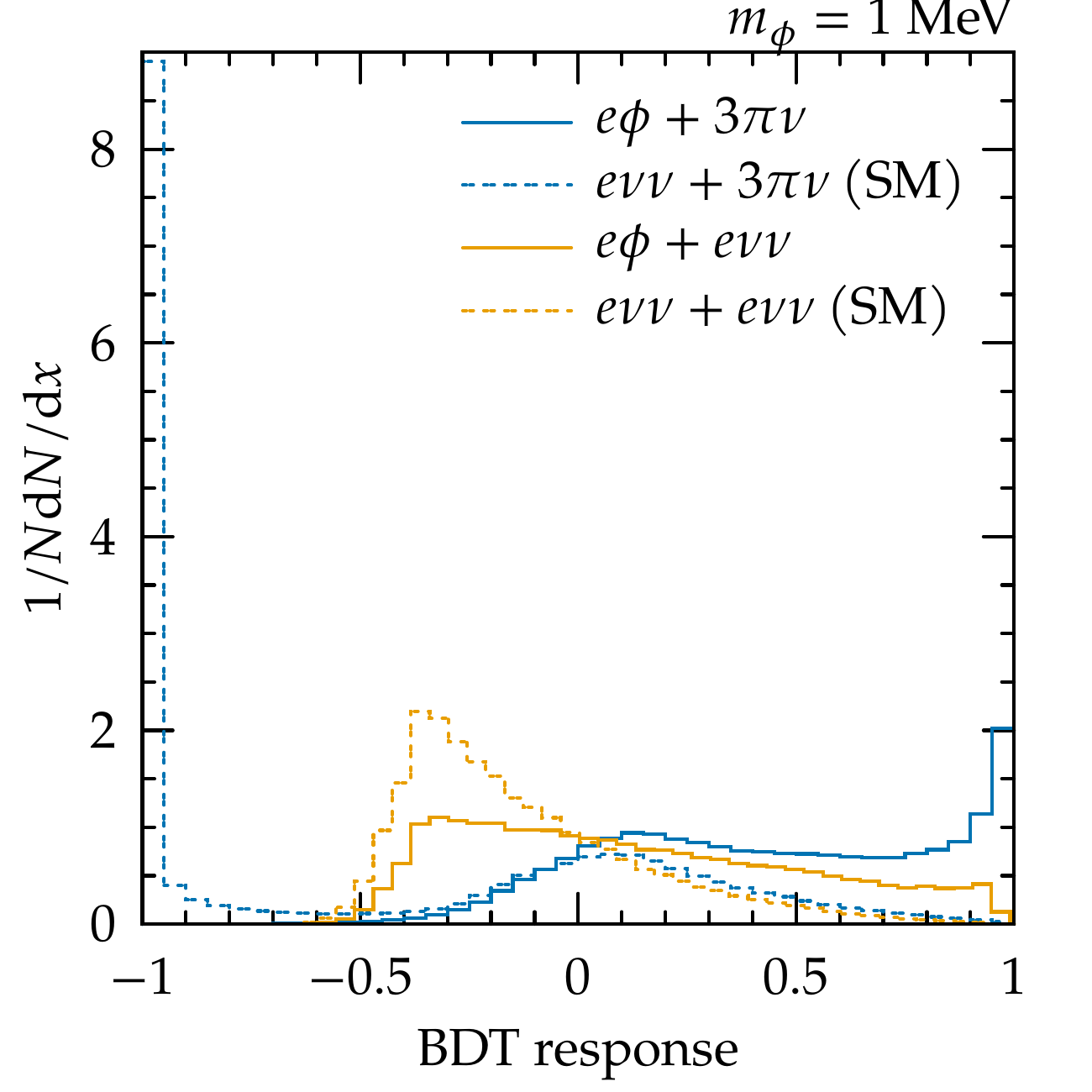}
    \includegraphics[width=0.30\textwidth]{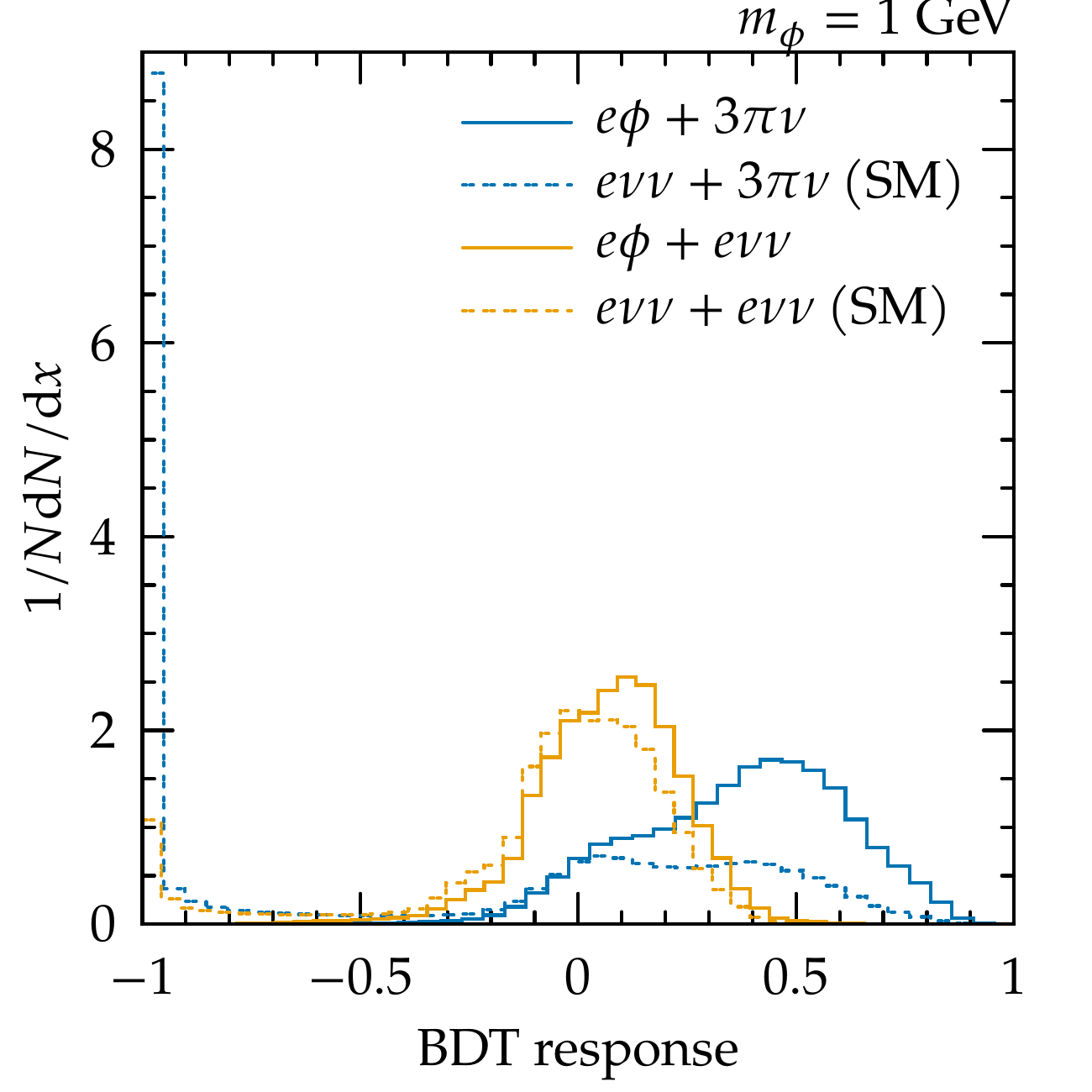}
  \end{center}
  \caption{\label{fig:BDT_ROC1GeV}
First panel: ROC curve for the case $\ma = 1~\GeV$, to be compared with the $1$-MeV case, shown in the first panel of fig. \ref{fig:ROC_UL} in the main text. Second and third panels: distributions in the BDT response, for both cases $\ma = 1~\MeV$ and $1$ GeV.
}
\end{figure*}

\bibliography{bibliography}

\begin{thebibliography}{37}%
\makeatletter
\providecommand \@ifxundefined [1]{%
 \@ifx{#1\undefined}
}%
\providecommand \@ifnum [1]{%
 \ifnum #1\expandafter \@firstoftwo
 \else \expandafter \@secondoftwo
 \fi
}%
\providecommand \@ifx [1]{%
 \ifx #1\expandafter \@firstoftwo
 \else \expandafter \@secondoftwo
 \fi
}%
\providecommand \natexlab [1]{#1}%
\providecommand \enquote  [1]{``#1''}%
\providecommand \bibnamefont  [1]{#1}%
\providecommand \bibfnamefont [1]{#1}%
\providecommand \citenamefont [1]{#1}%
\providecommand \href@noop [0]{\@secondoftwo}%
\providecommand \href [0]{\begingroup \@sanitize@url \@href}%
\providecommand \@href[1]{\@@startlink{#1}\@@href}%
\providecommand \@@href[1]{\endgroup#1\@@endlink}%
\providecommand \@sanitize@url [0]{\catcode `\\12\catcode `\$12\catcode
  `\&12\catcode `\#12\catcode `\^12\catcode `\_12\catcode `\%12\relax}%
\providecommand \@@startlink[1]{}%
\providecommand \@@endlink[0]{}%
\providecommand \url  [0]{\begingroup\@sanitize@url \@url }%
\providecommand \@url [1]{\endgroup\@href {#1}{\urlprefix }}%
\providecommand \urlprefix  [0]{URL }%
\providecommand \Eprint [0]{\href }%
\providecommand \doibase [0]{https://doi.org/}%
\providecommand \selectlanguage [0]{\@gobble}%
\providecommand \bibinfo  [0]{\@secondoftwo}%
\providecommand \bibfield  [0]{\@secondoftwo}%
\providecommand \translation [1]{[#1]}%
\providecommand \BibitemOpen [0]{}%
\providecommand \bibitemStop [0]{}%
\providecommand \bibitemNoStop [0]{.\EOS\space}%
\providecommand \EOS [0]{\spacefactor3000\relax}%
\providecommand \BibitemShut  [1]{\csname bibitem#1\endcsname}%
\let\auto@bib@innerbib\@empty
\bibitem [{\citenamefont {Beacham}\ \emph {et~al.}(2020)\citenamefont {Beacham}
  \emph {et~al.}}]{Beacham:2019nyx}%
  \BibitemOpen
  \bibfield  {author} {\bibinfo {author} {\bibfnamefont {J.}~\bibnamefont
  {Beacham}} \emph {et~al.},\ }\bibfield  {title} {\bibinfo {title} {{Physics
  Beyond Colliders at CERN: Beyond the Standard Model Working Group Report}},\
  }\href {https://doi.org/10.1088/1361-6471/ab4cd2} {\bibfield  {journal}
  {\bibinfo  {journal} {J. Phys.}\ }\textbf {\bibinfo {volume} {G47}},\
  \bibinfo {pages} {010501} (\bibinfo {year} {2020})},\ \Eprint
  {https://arxiv.org/abs/1901.09966} {arXiv:1901.09966 [hep-ex]} \BibitemShut
  {NoStop}%
\bibitem [{\citenamefont {Lanfranchi}\ \emph {et~al.}(2020)\citenamefont
  {Lanfranchi}, \citenamefont {Pospelov},\ and\ \citenamefont
  {Schuster}}]{Lanfranchi:2020crw}%
  \BibitemOpen
  \bibfield  {author} {\bibinfo {author} {\bibfnamefont {G.}~\bibnamefont
  {Lanfranchi}}, \bibinfo {author} {\bibfnamefont {M.}~\bibnamefont
  {Pospelov}},\ and\ \bibinfo {author} {\bibfnamefont {P.}~\bibnamefont
  {Schuster}},\ }\bibfield  {title} {\bibinfo {title} {{The Search for
  Feebly-Interacting Particles}}\ }\href
  {https://doi.org/10.1146/annurev-nucl-102419-055056}
  {10.1146/annurev-nucl-102419-055056} (\bibinfo {year} {2020}),\ \Eprint
  {https://arxiv.org/abs/2011.02157} {arXiv:2011.02157 [hep-ph]} \BibitemShut
  {NoStop}%
\bibitem [{\citenamefont {Georgi}\ \emph {et~al.}(1986)\citenamefont {Georgi},
  \citenamefont {Kaplan},\ and\ \citenamefont {Randall}}]{Georgi:1986df}%
  \BibitemOpen
  \bibfield  {author} {\bibinfo {author} {\bibfnamefont {H.}~\bibnamefont
  {Georgi}}, \bibinfo {author} {\bibfnamefont {D.~B.}\ \bibnamefont {Kaplan}},\
  and\ \bibinfo {author} {\bibfnamefont {L.}~\bibnamefont {Randall}},\
  }\bibfield  {title} {\bibinfo {title} {{Manifesting the Invisible Axion at
  Low-energies}},\ }\href {https://doi.org/10.1016/0370-2693(86)90688-X}
  {\bibfield  {journal} {\bibinfo  {journal} {Phys. Lett.}\ }\textbf {\bibinfo
  {volume} {169B}},\ \bibinfo {pages} {73} (\bibinfo {year}
  {1986})}\BibitemShut {NoStop}%
\bibitem [{\citenamefont {Calibbi}\ \emph {et~al.}(2020)\citenamefont
  {Calibbi}, \citenamefont {Redigolo}, \citenamefont {Ziegler},\ and\
  \citenamefont {Zupan}}]{Calibbi:2020jvd}%
  \BibitemOpen
  \bibfield  {author} {\bibinfo {author} {\bibfnamefont {L.}~\bibnamefont
  {Calibbi}}, \bibinfo {author} {\bibfnamefont {D.}~\bibnamefont {Redigolo}},
  \bibinfo {author} {\bibfnamefont {R.}~\bibnamefont {Ziegler}},\ and\ \bibinfo
  {author} {\bibfnamefont {J.}~\bibnamefont {Zupan}},\ }\bibfield  {title}
  {\bibinfo {title} {{Looking forward to Lepton-flavor-violating ALPs}},\
  }\href@noop {} {\  (\bibinfo {year} {2020})},\ \Eprint
  {https://arxiv.org/abs/2006.04795} {arXiv:2006.04795 [hep-ph]} \BibitemShut
  {NoStop}%
\bibitem [{\citenamefont {Baltrusaitis}\ \emph {et~al.}(1985)\citenamefont
  {Baltrusaitis} \emph {et~al.}}]{Baltrusaitis:1985fh}%
  \BibitemOpen
  \bibfield  {author} {\bibinfo {author} {\bibfnamefont {R.~M.}\ \bibnamefont
  {Baltrusaitis}} \emph {et~al.} (\bibinfo {collaboration} {MARK-III}),\
  }\bibfield  {title} {\bibinfo {title} {{$\tau$ Leptonic Branching Ratios and
  a Search for Goldstone Decay}},\ }\href
  {https://doi.org/10.1103/PhysRevLett.55.1842} {\bibfield  {journal} {\bibinfo
   {journal} {Phys. Rev. Lett.}\ }\textbf {\bibinfo {volume} {55}},\ \bibinfo
  {pages} {1842} (\bibinfo {year} {1985})}\BibitemShut {NoStop}%
\bibitem [{\citenamefont {Albrecht}\ \emph {et~al.}(1995)\citenamefont
  {Albrecht} \emph {et~al.}}]{Albrecht:1995ht}%
  \BibitemOpen
  \bibfield  {author} {\bibinfo {author} {\bibfnamefont {H.}~\bibnamefont
  {Albrecht}} \emph {et~al.} (\bibinfo {collaboration} {ARGUS}),\ }\bibfield
  {title} {\bibinfo {title} {{A Search for lepton flavor violating decays tau
  ----> e alpha, tau ---> mu alpha}},\ }\href
  {https://doi.org/10.1007/BF01579801} {\bibfield  {journal} {\bibinfo
  {journal} {Z. Phys.}\ }\textbf {\bibinfo {volume} {C68}},\ \bibinfo {pages}
  {25} (\bibinfo {year} {1995})}\BibitemShut {NoStop}%
\bibitem [{\citenamefont {Tenchini}\ \emph {et~al.}(2021)\citenamefont
  {Tenchini}, \citenamefont {Garcia-Hernandez}, \citenamefont {Kraetzschmar},
  \citenamefont {Rados}, \citenamefont {De~La Cruz-Burelo}, \citenamefont
  {De~Yta-Hernandez}, \citenamefont {Heredia de~la Cruz},\ and\ \citenamefont
  {Rostomyan}}]{Tenchini:2020njf}%
  \BibitemOpen
  \bibfield  {author} {\bibinfo {author} {\bibfnamefont {F.}~\bibnamefont
  {Tenchini}}, \bibinfo {author} {\bibfnamefont {M.}~\bibnamefont
  {Garcia-Hernandez}}, \bibinfo {author} {\bibfnamefont {T.}~\bibnamefont
  {Kraetzschmar}}, \bibinfo {author} {\bibfnamefont {P.~K.}\ \bibnamefont
  {Rados}}, \bibinfo {author} {\bibfnamefont {E.}~\bibnamefont {De~La
  Cruz-Burelo}}, \bibinfo {author} {\bibfnamefont {A.}~\bibnamefont
  {De~Yta-Hernandez}}, \bibinfo {author} {\bibfnamefont {I.}~\bibnamefont
  {Heredia de~la Cruz}},\ and\ \bibinfo {author} {\bibfnamefont
  {A.}~\bibnamefont {Rostomyan}},\ }\bibfield  {title} {\bibinfo {title}
  {{First results and prospects for tau LFV decay $\tau \rightarrow e +
  \alpha$(invisible) at Belle II}},\ }\bibfield  {booktitle} {\emph {\bibinfo
  {booktitle} {{Proceedings, 40th International Conference on High Energy
  Physics (ICHEP2020): Prague, Czechia, July 28 - August 6, 2020}}},\ }\href
  {https://doi.org/10.22323/1.390.0288} {\bibfield  {journal} {\bibinfo
  {journal} {PoS}\ }\textbf {\bibinfo {volume} {ICHEP2020}},\ \bibinfo {pages}
  {288} (\bibinfo {year} {2021})}\BibitemShut {NoStop}%
\bibitem [{\citenamefont {Bryman}\ \emph {et~al.}(2021)\citenamefont {Bryman},
  \citenamefont {Ito},\ and\ \citenamefont {Shrock}}]{Bryman:2021rtr}%
  \BibitemOpen
  \bibfield  {author} {\bibinfo {author} {\bibfnamefont {D.~A.}\ \bibnamefont
  {Bryman}}, \bibinfo {author} {\bibfnamefont {S.}~\bibnamefont {Ito}},\ and\
  \bibinfo {author} {\bibfnamefont {R.}~\bibnamefont {Shrock}},\ }\bibfield
  {title} {\bibinfo {title} {{Upper Limits on Branching Ratios of Decays $\tau
  \to \ell \gamma\gamma$ and $\tau \to \ell X$}},\ }\href@noop {} {\  (\bibinfo
  {year} {2021})},\ \Eprint {https://arxiv.org/abs/2106.02451}
  {arXiv:2106.02451 [hep-ph]} \BibitemShut {NoStop}%
\bibitem [{\citenamefont {Abudinén}\ \emph {et~al.}(2020)\citenamefont
  {Abudinén} \emph {et~al.}}]{Abudinen:2020das}%
  \BibitemOpen
  \bibfield  {author} {\bibinfo {author} {\bibfnamefont {F.}~\bibnamefont
  {Abudinén}} \emph {et~al.} (\bibinfo {collaboration} {Belle-II}),\
  }\bibfield  {title} {\bibinfo {title} {{$\tau$ lepton mass measurement at
  Belle II}},\ }\href@noop {} {\  (\bibinfo {year} {2020})},\ \Eprint
  {https://arxiv.org/abs/2008.04665} {arXiv:2008.04665 [hep-ex]} \BibitemShut
  {NoStop}%
\bibitem [{\citenamefont {Brandt}\ \emph {et~al.}(1964)\citenamefont {Brandt},
  \citenamefont {Peyrou}, \citenamefont {Sosnowski},\ and\ \citenamefont
  {Wroblewski}}]{Brandt:1964sa}%
  \BibitemOpen
  \bibfield  {author} {\bibinfo {author} {\bibfnamefont {S.}~\bibnamefont
  {Brandt}}, \bibinfo {author} {\bibfnamefont {C.}~\bibnamefont {Peyrou}},
  \bibinfo {author} {\bibfnamefont {R.}~\bibnamefont {Sosnowski}},\ and\
  \bibinfo {author} {\bibfnamefont {A.}~\bibnamefont {Wroblewski}},\ }\bibfield
   {title} {\bibinfo {title} {{The Principal axis of jets. An Attempt to
  analyze high-energy collisions as two-body processes}},\ }\href
  {https://doi.org/10.1016/0031-9163(64)91176-X} {\bibfield  {journal}
  {\bibinfo  {journal} {Phys. Lett.}\ }\textbf {\bibinfo {volume} {12}},\
  \bibinfo {pages} {57} (\bibinfo {year} {1964})}\BibitemShut {NoStop}%
\bibitem [{\citenamefont {Farhi}(1977)}]{Farhi:1977sg}%
  \BibitemOpen
  \bibfield  {author} {\bibinfo {author} {\bibfnamefont {E.}~\bibnamefont
  {Farhi}},\ }\bibfield  {title} {\bibinfo {title} {{A QCD Test for Jets}},\
  }\href {https://doi.org/10.1103/PhysRevLett.39.1587} {\bibfield  {journal}
  {\bibinfo  {journal} {Phys. Rev. Lett.}\ }\textbf {\bibinfo {volume} {39}},\
  \bibinfo {pages} {1587} (\bibinfo {year} {1977})}\BibitemShut {NoStop}%
\bibitem [{\citenamefont {Lester}\ and\ \citenamefont
  {Summers}(1999)}]{Lester:1999tx}%
  \BibitemOpen
  \bibfield  {author} {\bibinfo {author} {\bibfnamefont {C.~G.}\ \bibnamefont
  {Lester}}\ and\ \bibinfo {author} {\bibfnamefont {D.~J.}\ \bibnamefont
  {Summers}},\ }\bibfield  {title} {\bibinfo {title} {{Measuring masses of
  semiinvisibly decaying particles pair produced at hadron colliders}},\ }\href
  {https://doi.org/10.1016/S0370-2693(99)00945-4} {\bibfield  {journal}
  {\bibinfo  {journal} {Phys. Lett.}\ }\textbf {\bibinfo {volume} {B463}},\
  \bibinfo {pages} {99} (\bibinfo {year} {1999})},\ \Eprint
  {https://arxiv.org/abs/hep-ph/9906349} {arXiv:hep-ph/9906349 [hep-ph]}
  \BibitemShut {NoStop}%
\bibitem [{\citenamefont {Barr}\ \emph {et~al.}(2003)\citenamefont {Barr},
  \citenamefont {Lester},\ and\ \citenamefont {Stephens}}]{Barr:2003rg}%
  \BibitemOpen
  \bibfield  {author} {\bibinfo {author} {\bibfnamefont {A.}~\bibnamefont
  {Barr}}, \bibinfo {author} {\bibfnamefont {C.}~\bibnamefont {Lester}},\ and\
  \bibinfo {author} {\bibfnamefont {P.}~\bibnamefont {Stephens}},\ }\bibfield
  {title} {\bibinfo {title} {{m(T2): The Truth behind the glamour}},\ }\href
  {https://doi.org/10.1088/0954-3899/29/10/304} {\bibfield  {journal} {\bibinfo
   {journal} {J. Phys.}\ }\textbf {\bibinfo {volume} {G29}},\ \bibinfo {pages}
  {2343} (\bibinfo {year} {2003})},\ \Eprint
  {https://arxiv.org/abs/hep-ph/0304226} {arXiv:hep-ph/0304226 [hep-ph]}
  \BibitemShut {NoStop}%
\bibitem [{\citenamefont {Barr}\ \emph {et~al.}(2011)\citenamefont {Barr},
  \citenamefont {Khoo}, \citenamefont {Konar}, \citenamefont {Kong},
  \citenamefont {Lester}, \citenamefont {Matchev},\ and\ \citenamefont
  {Park}}]{Barr:2011xt}%
  \BibitemOpen
  \bibfield  {author} {\bibinfo {author} {\bibfnamefont {A.~J.}\ \bibnamefont
  {Barr}}, \bibinfo {author} {\bibfnamefont {T.~J.}\ \bibnamefont {Khoo}},
  \bibinfo {author} {\bibfnamefont {P.}~\bibnamefont {Konar}}, \bibinfo
  {author} {\bibfnamefont {K.}~\bibnamefont {Kong}}, \bibinfo {author}
  {\bibfnamefont {C.~G.}\ \bibnamefont {Lester}}, \bibinfo {author}
  {\bibfnamefont {K.~T.}\ \bibnamefont {Matchev}},\ and\ \bibinfo {author}
  {\bibfnamefont {M.}~\bibnamefont {Park}},\ }\bibfield  {title} {\bibinfo
  {title} {{Guide to transverse projections and mass-constraining variables}},\
  }\href {https://doi.org/10.1103/PhysRevD.84.095031} {\bibfield  {journal}
  {\bibinfo  {journal} {Phys. Rev.}\ }\textbf {\bibinfo {volume} {D84}},\
  \bibinfo {pages} {095031} (\bibinfo {year} {2011})},\ \Eprint
  {https://arxiv.org/abs/1105.2977} {arXiv:1105.2977 [hep-ph]} \BibitemShut
  {NoStop}%
\bibitem [{\citenamefont {Cho}\ \emph {et~al.}(2009)\citenamefont {Cho},
  \citenamefont {Choi}, \citenamefont {Kim},\ and\ \citenamefont
  {Park}}]{Cho:2008tj}%
  \BibitemOpen
  \bibfield  {author} {\bibinfo {author} {\bibfnamefont {W.~S.}\ \bibnamefont
  {Cho}}, \bibinfo {author} {\bibfnamefont {K.}~\bibnamefont {Choi}}, \bibinfo
  {author} {\bibfnamefont {Y.~G.}\ \bibnamefont {Kim}},\ and\ \bibinfo {author}
  {\bibfnamefont {C.~B.}\ \bibnamefont {Park}},\ }\bibfield  {title} {\bibinfo
  {title} {{M(T2)-assisted on-shell reconstruction of missing momenta and its
  application to spin measurement at the LHC}},\ }\href
  {https://doi.org/10.1103/PhysRevD.79.031701} {\bibfield  {journal} {\bibinfo
  {journal} {Phys. Rev.}\ }\textbf {\bibinfo {volume} {D79}},\ \bibinfo {pages}
  {031701} (\bibinfo {year} {2009})},\ \Eprint
  {https://arxiv.org/abs/0810.4853} {arXiv:0810.4853 [hep-ph]} \BibitemShut
  {NoStop}%
\bibitem [{\citenamefont {Park}(2011)}]{Park:2011uz}%
  \BibitemOpen
  \bibfield  {author} {\bibinfo {author} {\bibfnamefont {C.~B.}\ \bibnamefont
  {Park}},\ }\bibfield  {title} {\bibinfo {title} {{Reconstructing the heavy
  resonance at hadron colliders}},\ }\href
  {https://doi.org/10.1103/PhysRevD.84.096001} {\bibfield  {journal} {\bibinfo
  {journal} {Phys. Rev.}\ }\textbf {\bibinfo {volume} {D84}},\ \bibinfo {pages}
  {096001} (\bibinfo {year} {2011})},\ \Eprint
  {https://arxiv.org/abs/1106.6087} {arXiv:1106.6087 [hep-ph]} \BibitemShut
  {NoStop}%
\bibitem [{\citenamefont {Smith}\ \emph {et~al.}(1983)\citenamefont {Smith},
  \citenamefont {van Neerven},\ and\ \citenamefont
  {Vermaseren}}]{Smith:1983aa}%
  \BibitemOpen
  \bibfield  {author} {\bibinfo {author} {\bibfnamefont {J.}~\bibnamefont
  {Smith}}, \bibinfo {author} {\bibfnamefont {W.~L.}\ \bibnamefont {van
  Neerven}},\ and\ \bibinfo {author} {\bibfnamefont {J.~A.~M.}\ \bibnamefont
  {Vermaseren}},\ }\bibfield  {title} {\bibinfo {title} {{The Transverse Mass
  and Width of the $W$ Boson}},\ }\href
  {https://doi.org/10.1103/PhysRevLett.50.1738} {\bibfield  {journal} {\bibinfo
   {journal} {Phys. Rev. Lett.}\ }\textbf {\bibinfo {volume} {50}},\ \bibinfo
  {pages} {1738} (\bibinfo {year} {1983})}\BibitemShut {NoStop}%
\bibitem [{\citenamefont {Barger}\ \emph {et~al.}(1983)\citenamefont {Barger},
  \citenamefont {Martin},\ and\ \citenamefont {Phillips}}]{Barger:1983wf}%
  \BibitemOpen
  \bibfield  {author} {\bibinfo {author} {\bibfnamefont {V.~D.}\ \bibnamefont
  {Barger}}, \bibinfo {author} {\bibfnamefont {A.~D.}\ \bibnamefont {Martin}},\
  and\ \bibinfo {author} {\bibfnamefont {R.~J.~N.}\ \bibnamefont {Phillips}},\
  }\bibfield  {title} {\bibinfo {title} {{Perpendicular $\nu_e$ Mass From $W$
  Decay}},\ }\bibfield  {booktitle} {\emph {\bibinfo {booktitle} {{11th
  International Symposium on Lepton and Photon Interactions at High Energies
  Ithaca, New York, August 4-9, 1983}}},\ }\href
  {https://doi.org/10.1007/BF01648783} {\bibfield  {journal} {\bibinfo
  {journal} {Z. Phys.}\ }\textbf {\bibinfo {volume} {C21}},\ \bibinfo {pages}
  {99} (\bibinfo {year} {1983})}\BibitemShut {NoStop}%
\bibitem [{\citenamefont {Ross}\ and\ \citenamefont
  {Serna}(2008)}]{Ross:2007rm}%
  \BibitemOpen
  \bibfield  {author} {\bibinfo {author} {\bibfnamefont {G.~G.}\ \bibnamefont
  {Ross}}\ and\ \bibinfo {author} {\bibfnamefont {M.}~\bibnamefont {Serna}},\
  }\bibfield  {title} {\bibinfo {title} {{Mass determination of new states at
  hadron colliders}},\ }\href {https://doi.org/10.1016/j.physletb.2008.06.003}
  {\bibfield  {journal} {\bibinfo  {journal} {Phys. Lett.}\ }\textbf {\bibinfo
  {volume} {B665}},\ \bibinfo {pages} {212} (\bibinfo {year} {2008})},\ \Eprint
  {https://arxiv.org/abs/0712.0943} {arXiv:0712.0943 [hep-ph]} \BibitemShut
  {NoStop}%
\bibitem [{\citenamefont {Cho}\ \emph {et~al.}(2014)\citenamefont {Cho},
  \citenamefont {Gainer}, \citenamefont {Kim}, \citenamefont {Matchev},
  \citenamefont {Moortgat}, \citenamefont {Pape},\ and\ \citenamefont
  {Park}}]{Cho:2014naa}%
  \BibitemOpen
  \bibfield  {author} {\bibinfo {author} {\bibfnamefont {W.~S.}\ \bibnamefont
  {Cho}}, \bibinfo {author} {\bibfnamefont {J.~S.}\ \bibnamefont {Gainer}},
  \bibinfo {author} {\bibfnamefont {D.}~\bibnamefont {Kim}}, \bibinfo {author}
  {\bibfnamefont {K.~T.}\ \bibnamefont {Matchev}}, \bibinfo {author}
  {\bibfnamefont {F.}~\bibnamefont {Moortgat}}, \bibinfo {author}
  {\bibfnamefont {L.}~\bibnamefont {Pape}},\ and\ \bibinfo {author}
  {\bibfnamefont {M.}~\bibnamefont {Park}},\ }\bibfield  {title} {\bibinfo
  {title} {{On-shell constrained $M_2$ variables with applications to mass
  measurements and topology disambiguation}},\ }\href
  {https://doi.org/10.1007/JHEP08(2014)070} {\bibfield  {journal} {\bibinfo
  {journal} {JHEP}\ }\textbf {\bibinfo {volume} {08}},\ \bibinfo {pages}
  {070}},\ \Eprint {https://arxiv.org/abs/1401.1449} {arXiv:1401.1449 [hep-ph]}
  \BibitemShut {NoStop}%
\bibitem [{\citenamefont {Konar}\ and\ \citenamefont
  {Swain}(2016{\natexlab{a}})}]{Konar:2015hea}%
  \BibitemOpen
  \bibfield  {author} {\bibinfo {author} {\bibfnamefont {P.}~\bibnamefont
  {Konar}}\ and\ \bibinfo {author} {\bibfnamefont {A.~K.}\ \bibnamefont
  {Swain}},\ }\bibfield  {title} {\bibinfo {title} {{Mass reconstruction with
  $M_2$ under constraint in semi-invisible production at a hadron collider}},\
  }\href {https://doi.org/10.1103/PhysRevD.93.015021} {\bibfield  {journal}
  {\bibinfo  {journal} {Phys. Rev.}\ }\textbf {\bibinfo {volume} {D93}},\
  \bibinfo {pages} {015021} (\bibinfo {year} {2016}{\natexlab{a}})},\ \Eprint
  {https://arxiv.org/abs/1509.00298} {arXiv:1509.00298 [hep-ph]} \BibitemShut
  {NoStop}%
\bibitem [{\citenamefont {Konar}\ and\ \citenamefont
  {Swain}(2016{\natexlab{b}})}]{Konar:2016wbh}%
  \BibitemOpen
  \bibfield  {author} {\bibinfo {author} {\bibfnamefont {P.}~\bibnamefont
  {Konar}}\ and\ \bibinfo {author} {\bibfnamefont {A.~K.}\ \bibnamefont
  {Swain}},\ }\bibfield  {title} {\bibinfo {title} {{Reconstructing
  semi-invisible events in resonant tau pair production from Higgs}},\ }\href
  {https://doi.org/10.1016/j.physletb.2016.03.070} {\bibfield  {journal}
  {\bibinfo  {journal} {Phys. Lett.}\ }\textbf {\bibinfo {volume} {B757}},\
  \bibinfo {pages} {211} (\bibinfo {year} {2016}{\natexlab{b}})},\ \Eprint
  {https://arxiv.org/abs/1602.00552} {arXiv:1602.00552 [hep-ph]} \BibitemShut
  {NoStop}%
\bibitem [{\citenamefont {Agashe}\ \emph {et~al.}(2011)\citenamefont {Agashe},
  \citenamefont {Kim}, \citenamefont {Walker},\ and\ \citenamefont
  {Zhu}}]{Agashe:2010tu}%
  \BibitemOpen
  \bibfield  {author} {\bibinfo {author} {\bibfnamefont {K.}~\bibnamefont
  {Agashe}}, \bibinfo {author} {\bibfnamefont {D.}~\bibnamefont {Kim}},
  \bibinfo {author} {\bibfnamefont {D.~G.~E.}\ \bibnamefont {Walker}},\ and\
  \bibinfo {author} {\bibfnamefont {L.}~\bibnamefont {Zhu}},\ }\bibfield
  {title} {\bibinfo {title} {{Using $M_{T2}$ to Distinguish Dark Matter
  Stabilization Symmetries}},\ }\href
  {https://doi.org/10.1103/PhysRevD.84.055020} {\bibfield  {journal} {\bibinfo
  {journal} {Phys. Rev.}\ }\textbf {\bibinfo {volume} {D84}},\ \bibinfo {pages}
  {055020} (\bibinfo {year} {2011})},\ \Eprint
  {https://arxiv.org/abs/1012.4460} {arXiv:1012.4460 [hep-ph]} \BibitemShut
  {NoStop}%
\bibitem [{\citenamefont {Giudice}\ \emph {et~al.}(2012)\citenamefont
  {Giudice}, \citenamefont {Gripaios},\ and\ \citenamefont
  {Mahbubani}}]{Giudice:2011ib}%
  \BibitemOpen
  \bibfield  {author} {\bibinfo {author} {\bibfnamefont {G.~F.}\ \bibnamefont
  {Giudice}}, \bibinfo {author} {\bibfnamefont {B.}~\bibnamefont {Gripaios}},\
  and\ \bibinfo {author} {\bibfnamefont {R.}~\bibnamefont {Mahbubani}},\
  }\bibfield  {title} {\bibinfo {title} {{Counting dark matter particles in LHC
  events}},\ }\href {https://doi.org/10.1103/PhysRevD.85.075019} {\bibfield
  {journal} {\bibinfo  {journal} {Phys. Rev.}\ }\textbf {\bibinfo {volume}
  {D85}},\ \bibinfo {pages} {075019} (\bibinfo {year} {2012})},\ \Eprint
  {https://arxiv.org/abs/1108.1800} {arXiv:1108.1800 [hep-ph]} \BibitemShut
  {NoStop}%
\bibitem [{\citenamefont {Kim}\ \emph {et~al.}(2017)\citenamefont {Kim},
  \citenamefont {Matchev}, \citenamefont {Moortgat},\ and\ \citenamefont
  {Pape}}]{Kim:2017awi}%
  \BibitemOpen
  \bibfield  {author} {\bibinfo {author} {\bibfnamefont {D.}~\bibnamefont
  {Kim}}, \bibinfo {author} {\bibfnamefont {K.~T.}\ \bibnamefont {Matchev}},
  \bibinfo {author} {\bibfnamefont {F.}~\bibnamefont {Moortgat}},\ and\
  \bibinfo {author} {\bibfnamefont {L.}~\bibnamefont {Pape}},\ }\bibfield
  {title} {\bibinfo {title} {{Testing Invisible Momentum Ansatze in Missing
  Energy Events at the LHC}},\ }\href {https://doi.org/10.1007/JHEP08(2017)102}
  {\bibfield  {journal} {\bibinfo  {journal} {JHEP}\ }\textbf {\bibinfo
  {volume} {08}},\ \bibinfo {pages} {102}},\ \Eprint
  {https://arxiv.org/abs/1703.06887} {arXiv:1703.06887 [hep-ph]} \BibitemShut
  {NoStop}%
\bibitem [{\citenamefont {Li}\ \emph {et~al.}(2012)\citenamefont {Li},
  \citenamefont {Ito}, \citenamefont {Poschl}, \citenamefont {Richard},
  \citenamefont {Ruan}, \citenamefont {Takubo},\ and\ \citenamefont
  {Yamamoto}}]{Li:2012taa}%
  \BibitemOpen
  \bibfield  {author} {\bibinfo {author} {\bibfnamefont {H.}~\bibnamefont
  {Li}}, \bibinfo {author} {\bibfnamefont {K.}~\bibnamefont {Ito}}, \bibinfo
  {author} {\bibfnamefont {R.}~\bibnamefont {Poschl}}, \bibinfo {author}
  {\bibfnamefont {F.}~\bibnamefont {Richard}}, \bibinfo {author} {\bibfnamefont
  {M.}~\bibnamefont {Ruan}}, \bibinfo {author} {\bibfnamefont {Y.}~\bibnamefont
  {Takubo}},\ and\ \bibinfo {author} {\bibfnamefont {H.}~\bibnamefont
  {Yamamoto}} (\bibinfo {collaboration} {ILD Design Study Group}),\ }\bibfield
  {title} {\bibinfo {title} {{HZ Recoil Mass and Cross Section Analysis in
  ILD}},\ }\href@noop {} {\  (\bibinfo {year} {2012})},\ \Eprint
  {https://arxiv.org/abs/1202.1439} {arXiv:1202.1439 [hep-ex]} \BibitemShut
  {NoStop}%
\bibitem [{\citenamefont {Fujii}\ \emph {et~al.}(2015)\citenamefont {Fujii}
  \emph {et~al.}}]{Fujii:2015jha}%
  \BibitemOpen
  \bibfield  {author} {\bibinfo {author} {\bibfnamefont {K.}~\bibnamefont
  {Fujii}} \emph {et~al.},\ }\bibfield  {title} {\bibinfo {title} {{Physics
  Case for the International Linear Collider}},\ }\href@noop {} {\  (\bibinfo
  {year} {2015})},\ \Eprint {https://arxiv.org/abs/1506.05992}
  {arXiv:1506.05992 [hep-ex]} \BibitemShut {NoStop}%
\bibitem [{\citenamefont {De~La Cruz-Burelo}\ \emph {et~al.}(2020)\citenamefont
  {De~La Cruz-Burelo}, \citenamefont {De~Yta-Hernandez},\ and\ \citenamefont
  {Hernandez-Villanueva}}]{DeLaCruz-Burelo:2020ozf}%
  \BibitemOpen
  \bibfield  {author} {\bibinfo {author} {\bibfnamefont {E.}~\bibnamefont
  {De~La Cruz-Burelo}}, \bibinfo {author} {\bibfnamefont {A.}~\bibnamefont
  {De~Yta-Hernandez}},\ and\ \bibinfo {author} {\bibfnamefont {M.}~\bibnamefont
  {Hernandez-Villanueva}},\ }\bibfield  {title} {\bibinfo {title} {{New method
  for beyond the Standard Model invisible particle searches in tau lepton
  decays}},\ }\href {https://doi.org/10.1103/PhysRevD.102.115001} {\bibfield
  {journal} {\bibinfo  {journal} {Phys. Rev.}\ }\textbf {\bibinfo {volume}
  {D102}},\ \bibinfo {pages} {115001} (\bibinfo {year} {2020})},\ \Eprint
  {https://arxiv.org/abs/2007.08239} {arXiv:2007.08239 [hep-ph]} \BibitemShut
  {NoStop}%
\bibitem [{\citenamefont {Hagiwara}\ \emph {et~al.}(2013)\citenamefont
  {Hagiwara}, \citenamefont {Li}, \citenamefont {Mawatari},\ and\ \citenamefont
  {Nakamura}}]{Hagiwara:2012vz}%
  \BibitemOpen
  \bibfield  {author} {\bibinfo {author} {\bibfnamefont {K.}~\bibnamefont
  {Hagiwara}}, \bibinfo {author} {\bibfnamefont {T.}~\bibnamefont {Li}},
  \bibinfo {author} {\bibfnamefont {K.}~\bibnamefont {Mawatari}},\ and\
  \bibinfo {author} {\bibfnamefont {J.}~\bibnamefont {Nakamura}},\ }\bibfield
  {title} {\bibinfo {title} {{TauDecay: a library to simulate polarized tau
  decays via FeynRules and MadGraph5}},\ }\href
  {https://doi.org/10.1140/epjc/s10052-013-2489-4} {\bibfield  {journal}
  {\bibinfo  {journal} {Eur. Phys. J.}\ }\textbf {\bibinfo {volume} {C73}},\
  \bibinfo {pages} {2489} (\bibinfo {year} {2013})},\ \Eprint
  {https://arxiv.org/abs/1212.6247} {arXiv:1212.6247 [hep-ph]} \BibitemShut
  {NoStop}%
\bibitem [{\citenamefont {Bertacchi}\ \emph {et~al.}(2021)\citenamefont
  {Bertacchi} \emph {et~al.}}]{BelleIITrackingGroup:2020hpx}%
  \BibitemOpen
  \bibfield  {author} {\bibinfo {author} {\bibfnamefont {V.}~\bibnamefont
  {Bertacchi}} \emph {et~al.} (\bibinfo {collaboration} {Belle II Tracking
  Group}),\ }\bibfield  {title} {\bibinfo {title} {{Track finding at Belle
  II}},\ }\href {https://doi.org/10.1016/j.cpc.2020.107610} {\bibfield
  {journal} {\bibinfo  {journal} {Comput. Phys. Commun.}\ }\textbf {\bibinfo
  {volume} {259}},\ \bibinfo {pages} {107610} (\bibinfo {year} {2021})},\
  \Eprint {https://arxiv.org/abs/2003.12466} {arXiv:2003.12466
  [physics.ins-det]} \BibitemShut {NoStop}%
\bibitem [{\citenamefont {Park}(2021)}]{Park:2020bsu}%
  \BibitemOpen
  \bibfield  {author} {\bibinfo {author} {\bibfnamefont {C.~B.}\ \bibnamefont
  {Park}},\ }\bibfield  {title} {\bibinfo {title} {{YAM2: Yet another library
  for the $M_2$ variables using sequential quadratic programming}},\ }\href
  {https://doi.org/10.1016/j.cpc.2021.107967} {\bibfield  {journal} {\bibinfo
  {journal} {Comput. Phys. Commun.}\ }\textbf {\bibinfo {volume} {264}},\
  \bibinfo {pages} {107967} (\bibinfo {year} {2021})},\ \Eprint
  {https://arxiv.org/abs/2007.15537} {arXiv:2007.15537 [hep-ph]} \BibitemShut
  {NoStop}%
\bibitem [{\citenamefont {Hocker}\ \emph {et~al.}(2007)\citenamefont {Hocker}
  \emph {et~al.}}]{Hocker:2007ht}%
  \BibitemOpen
  \bibfield  {author} {\bibinfo {author} {\bibfnamefont {A.}~\bibnamefont
  {Hocker}} \emph {et~al.},\ }\bibfield  {title} {\bibinfo {title} {{TMVA -
  Toolkit for Multivariate Data Analysis}},\ }\href@noop {} {\  (\bibinfo
  {year} {2007})},\ \Eprint {https://arxiv.org/abs/physics/0703039}
  {arXiv:physics/0703039} \BibitemShut {NoStop}%
\bibitem [{\citenamefont {Zyla}\ \emph {et~al.}(2020)\citenamefont {Zyla} \emph
  {et~al.}}]{Zyla:2020zbs}%
  \BibitemOpen
  \bibfield  {author} {\bibinfo {author} {\bibfnamefont {P.~A.}\ \bibnamefont
  {Zyla}} \emph {et~al.} (\bibinfo {collaboration} {Particle Data Group}),\
  }\bibfield  {title} {\bibinfo {title} {{Review of Particle Physics}},\ }\href
  {https://doi.org/10.1093/ptep/ptaa104} {\bibfield  {journal} {\bibinfo
  {journal} {PTEP}\ }\textbf {\bibinfo {volume} {2020}},\ \bibinfo {pages}
  {083C01} (\bibinfo {year} {2020})}\BibitemShut {NoStop}%
\bibitem [{\citenamefont {Guadagnoli}\ and\ \citenamefont
  {Park}(2014)}]{Guadagnoli:2013xia}%
  \BibitemOpen
  \bibfield  {author} {\bibinfo {author} {\bibfnamefont {D.}~\bibnamefont
  {Guadagnoli}}\ and\ \bibinfo {author} {\bibfnamefont {C.~B.}\ \bibnamefont
  {Park}},\ }\bibfield  {title} {\bibinfo {title} {{$M_{T2}$-reconstructed
  invisible momenta as spin analizers, and an application to top
  polarization}},\ }\href {https://doi.org/10.1007/JHEP01(2014)030} {\bibfield
  {journal} {\bibinfo  {journal} {JHEP}\ }\textbf {\bibinfo {volume} {01}},\
  \bibinfo {pages} {030}},\ \Eprint {https://arxiv.org/abs/1308.2226}
  {arXiv:1308.2226 [hep-ph]} \BibitemShut {NoStop}%
\bibitem [{\citenamefont {Choi}\ \emph {et~al.}(2011)\citenamefont {Choi},
  \citenamefont {Guadagnoli},\ and\ \citenamefont {Park}}]{Choi:2011ys}%
  \BibitemOpen
  \bibfield  {author} {\bibinfo {author} {\bibfnamefont {K.}~\bibnamefont
  {Choi}}, \bibinfo {author} {\bibfnamefont {D.}~\bibnamefont {Guadagnoli}},\
  and\ \bibinfo {author} {\bibfnamefont {C.~B.}\ \bibnamefont {Park}},\
  }\bibfield  {title} {\bibinfo {title} {{Reducing combinatorial uncertainties:
  A new technique based on MT2 variables}},\ }\href
  {https://doi.org/10.1007/JHEP11(2011)117} {\bibfield  {journal} {\bibinfo
  {journal} {JHEP}\ }\textbf {\bibinfo {volume} {11}},\ \bibinfo {pages}
  {117}},\ \Eprint {https://arxiv.org/abs/1109.2201} {arXiv:1109.2201 [hep-ph]}
  \BibitemShut {NoStop}%
\bibitem [{\citenamefont {Bhattiprolu}\ \emph {et~al.}(2021)\citenamefont
  {Bhattiprolu}, \citenamefont {Martin},\ and\ \citenamefont
  {Wells}}]{Bhattiprolu:2020mwi}%
  \BibitemOpen
  \bibfield  {author} {\bibinfo {author} {\bibfnamefont {P.~N.}\ \bibnamefont
  {Bhattiprolu}}, \bibinfo {author} {\bibfnamefont {S.~P.}\ \bibnamefont
  {Martin}},\ and\ \bibinfo {author} {\bibfnamefont {J.~D.}\ \bibnamefont
  {Wells}},\ }\bibfield  {title} {\bibinfo {title} {{Criteria for projected
  discovery and exclusion sensitivities of counting experiments}},\ }\href
  {https://doi.org/10.1140/epjc/s10052-020-08819-6} {\bibfield  {journal}
  {\bibinfo  {journal} {Eur. Phys. J.}\ }\textbf {\bibinfo {volume} {C81}},\
  \bibinfo {pages} {123} (\bibinfo {year} {2021})},\ \Eprint
  {https://arxiv.org/abs/2009.07249} {arXiv:2009.07249 [physics.data-an]}
  \BibitemShut {NoStop}%
\bibitem [{Red()}]{Redigolo:GDR}%
  \BibitemOpen
  \href@noop {} {}\bibinfo {note} {D.~Redigolo, ``{Linking axions/ALPs with
  flavour, and searches at flavour experiments},'', GDR Intensity Frontier
  Annual Meeting 2020, \newblock
  \url{https://indico.in2p3.fr/event/22193}.}\BibitemShut {Stop}%
\end{thebibliography}%

\end{document}